\documentclass[journal]{IEEEtran}

\usepackage{ifpdf}
\usepackage{cite}
\usepackage{amsmath,amssymb,amsfonts}
\usepackage{algorithmic}
\usepackage{graphicx, adjustbox}
\usepackage{array}
\usepackage{enumitem}
\usepackage{breqn}
\usepackage{ dsfont }
\usepackage{ upgreek }
\usepackage{textcomp}
\usepackage{multirow}
\usepackage{algorithm}
\usepackage{algorithmic}
\usepackage{caption}
\usepackage{pgfplots}
\usepackage{tkz-euclide,subfigure}
\usepackage{tikz}
\usepackage{hyperref}
\usepackage{balance}

\begin{document}
	
	\title{BEdgeHealth: A Decentralized Architecture for Edge-based IoMT Networks Using Blockchain}
	
	\author{Dinh C. Nguyen,~\IEEEmembership{Member,~IEEE,} Pubudu N. Pathirana,~\IEEEmembership{Senior Member,~IEEE,} \\
		Ming Ding,~\IEEEmembership{Senior Member,~IEEE,} and
		Aruna Seneviratne,~\IEEEmembership{Senior Member,~IEEE}
		\thanks {*Part of this work has been accepted at the IEEE Global Communications Conference (GLOBECOM), Taiwan, 2020 \cite{DinhGlobe}.}
		\thanks{Dinh C. Nguyen is with School of Engineering, Deakin University, Waurn Ponds, VIC 3216, Australia, and also with the Data61, CSIRO, Docklands, Melbourne, Australia  (e-mail: cdnguyen@deakin.edu.au).}
		\thanks{Pubudu N. Pathirana is with School of Engineering, Deakin University, Waurn Ponds, VIC 3216, Australia (email: pubudu.pathirana@deakin.edu.au).}
		\thanks{Ming Ding is with the Data61, CSIRO, Australia (email: ming.ding@data61.csiro.au).}
		\thanks{Aruna Seneviratne is with School of Electrical Engineering and Telecommunications, University of New South Wales (UNSW), NSW, Australia (email: a.seneviratne@unsw.edu.au).}
	}
	
	\markboth{Accepted at IEEE Internet of Things Journal}%
	{}

	\maketitle
	
\begin{abstract}
	The healthcare industry has witnessed significant transformations in e-health services by using mobile edge computing (MEC) and blockchain to facilitate healthcare operations. Many MEC-blockchain-based schemes have been proposed, but some critical technical challenges still remain, such as low quality of services (QoS), data privacy and system security vulnerabilities. In this paper, we propose a new decentralized health architecture, called \textit{BEdgeHealth} that integrates MEC and blockchain for data offloading and data sharing in distributed hospital networks. First, a data offloading scheme is proposed where mobile devices can offload health data to a nearby MEC server for efficient computation with privacy awareness. Moreover, we design a data sharing scheme which enables data exchanges among healthcare users by leveraging blockchain and interplanetary file system. Particularly, a smart contract-based authentication mechanism is integrated with MEC to perform decentralized user access verification at the network edge without requiring any central authority. The real-world experiment results and evaluations demonstrate the effectiveness of the proposed \textit{BEdgeHealth} architecture in terms of \textcolor{black}{improved QoS with data privacy and security guarantees}, compared to the existing schemes. 
\end{abstract}


\IEEEpeerreviewmaketitle

\section{Introduction}

Recent advances in mobile edge computing (MEC) and Internet of Medical Things (IoMT) technologies have promoted smart e-health services \cite{3} in the healthcare industry. In hospital networks, health data collected from mobile devices (MDs), i.e., smartphones, laptops, and tablets, can be offloaded to nearby MEC servers for low-latency data execution, which thus helps improve users' computation experience, enhance quality of services (QoS) and reduce computing burden on local devices. Further, MEC also enables the ubiquitous sharing of health data acquired from the offloading phase among health users  to support healthcare delivery \cite{2}. For example, a doctor can exploit health data at the network edge to serve disease diagnosis and treatment, and patients can gain medical benefits such as medication advice in a low-latency fashion. Particularly, blockchain, a disruptive technology for enabling healthcare in recent years, is able to provide high security for health data sharing due to its unique features such as decentralization, traceability, and immutability \cite{mcghin2019blockchain}. In fact, blockchain can ensure reliable data exchange among healthcare users such as healthcare providers, insurance companies, and patients in smart healthcare environments like cooperative hospital networks \cite{yazdinejad2020decentralized}, by using immutable data ledgers and smart contracts \cite{mcghin2019blockchain}. Enabled by the computing efficiency of MEC and the security features of blockchain, the combination of these two emerging technologies has been regarded as the key enabler for smart healthcare services, i.e., data offloading and data sharing, in distributed hospital settings \cite{3}.

However, realizing the potential of such a comprehensive system in healthcare still faces non-trivial challenges. First, how to offload IoT healthcare data to the MEC servers to support e-health applications while guaranteeing both high QoS (i.e., minimal service latency) and data privacy is a critical issue. Most traditional approaches \cite{4,5} only either focus on the QoS problem of network latency and energy usage or data privacy for the healthcare offloading, while building a holistic framework with all these factors taken into consideration is vitally necessary. Second, some data sharing solutions in \cite{11} rely on a centralized cloud architectur which is prone to single-point failures and raises trust concerns from third party. The storage of electronic health records (EHRs) in central cloud also incurs high communication overhead, although it requires less data management efforts. Final, some blockchain-based proposals for health data sharing have been introduced in recent works \cite{13,14,30}. \textcolor{black}{However, these schemes often use a classic interplanetary file system (IPFS) which relies on a global Distributed Hash Table (DHT) for health data storage and sharing, leading to high data retrieval latency. To be clear, to access a record stored on an IPFS node, one needs to refer to a DHT to obtain the hash and then returns to the IPFS node for retrieving data, which incurs high data sharing overhead.} Moreover, the potential of blockchain for healthcare sharing scenarios with MEC in cooperative hospital networks has not been adequately investigated. 

\textcolor{black}{To fill these research gaps, this paper proposes a new decentralized health architecture which integrates a data offloading scheme and a data sharing scheme for distributed hospital networks with MEC and blockchain. Particularly, we design  a new decentralized smart contract associated with IPFS running on top of the MEC network which brings two key benefits. First, the smart contract is able to provide authentication and traceability in the data sharing \cite{13}. Any upload events or user access behaviours in the hospital network will be authenticated and traced by the contract without the need of external authority. Further, any modification or alternation on the data record will lead to a change on hash value, which can be also traced by the contract. Second, the combination of smart contract and IPFS helps accelerate the data retrieval rate. Here, we make an improvement in IPFS design by eliminating the global DHT. Instead, we store directly the hash values of data records in the smart contract at each of the MEC servers, which improves the data lookup and data retrieval rates in the data sharing.}

\subsection{Motivations and Contributions}

First, the work in \cite{8} makes us realize that the research on health data offloading and data sharing is of great practical significance in facilitating smart healthcare. Second, the existing studies on mobile health data offloading \cite{9}, \cite{10} motivate us to focus on the MEC framework for highly efficient health data computation. Moreover, the analysis in \cite{12}, \cite{13} and the preliminary results from our recent work \cite{14} reveal that blockchain can provide promising health data sharing solutions with improved QoS and enhanced security. Third, the in-depth discussion in \cite{mcghin2019blockchain} highlights the urgent need of developing a comprehensive architecture of data offloading and data sharing for improving the healthcare quality. Therefore, the above existing works strengthen our determination to implement a comprehensive healthcare architecture by using blockchain, IPFS, smart contract and MEC technologies. The key contributions of this paper are summarized as follows:
\begin{enumerate}
	\item 	We propose a new \textit{decentralized} health architecture for a cooperative hospital network, called \textit{BEdgeHealth} that integrates blockchain and MEC for data offloading and data sharing with \textcolor{black}{user QoS and security awareness}.  
	\item 	We propose a privacy-aware health data offloading scheme where MDs can offload data tasks to a nearby MEC server for efficient computation. An optimization algorithm is built on each MD to minimize the offloading cost of energy consumption, processing time, and memory usage with respect to system constraints.
	\item  We develop a data sharing scheme enabled by the cooperation of blockchain, MEC, smart contract and IPFS in the distributed hospital network. Particularly, a decentralized authentication mechanism associated with a distributed IPFS storage is built to implement access control and data management at the network edge without requiring third party, in order to enhance the sharing security and improve data retrieval rate.  
	\item We conduct real-world experiments to evaluate the effectiveness of the proposed \textit{BEdgeHealth} architecture. The implementation results and discussions demonstrate the superior performance of our proposed approach over the existing works.
	
\end{enumerate}
\subsection{Organization}
The remainder of the paper is organized as follows. Section~\ref {Relatedwork} discusses the literature works related to health data offloading and data sharing. In Section~\ref{SystemModel}, we introduce our decentralized health architecture where the network components and blockchain are explained. We then present our data offloading scheme in Section~\ref{Offloading} where the offloading model is formulated. Next,  Section~\ref{Sharing} introduces the health data sharing scheme using blockchain in distributed hospitals. The experimental results are provided in Section~\ref{Experimental}, while Section~\ref{Security} presents the security analysis and discussions. Finally, Section~\ref{Conclude} concludes the paper and highlights some possible future works.
\section{Related Works}
\label{Relatedwork}
In this section, we survey the literature works related to health data offloading and sharing in healthcare.
\subsection{Health Data Offloading}
Many data offloading approaches have been proposed to support healthcare. In \cite{4}, mobile healthcare data can be offloaded to cloud for processing, analysis, and storage, but it remains high latency incurred by offloading data to remote clouds. Also, the offloading privacy is not considered, which puts sensitive health data at risks of external attacks. Another work in \cite{8} proposed an IoT architecture for executing healthcare applications on clouds, but the optimization for memory usage of MDs in the offloading has been overlooked. In fact, there is close relationship between memory usage and the data task as investigated in the recent work \cite{add1}. That is, the higher size of data tasks to be offloaded, the higher device memory required to handle the task. Therefore, it is important to consider device memory usage in the data task offloading. Other works in \cite{5,9,10} concentrated on offloading security issues in healthcare. For example, the research in \cite{9} used a hash function and a key cryptosystem for data security. Also, the offloading privacy issues were also solved in \cite{5}, \cite{10} by using consensus algorithms and learning-based privacy preservation techniques with respect to response time and delay requirements. However, the above studies lack the joint consideration of all QoS constraints (i.e., network latency, energy consumption and memory usage) and privacy awareness, which are the important aspects of practical health data offloading systems \cite{2}. 
\begin{figure*}
	\centering
	\includegraphics[width=0.95\linewidth]{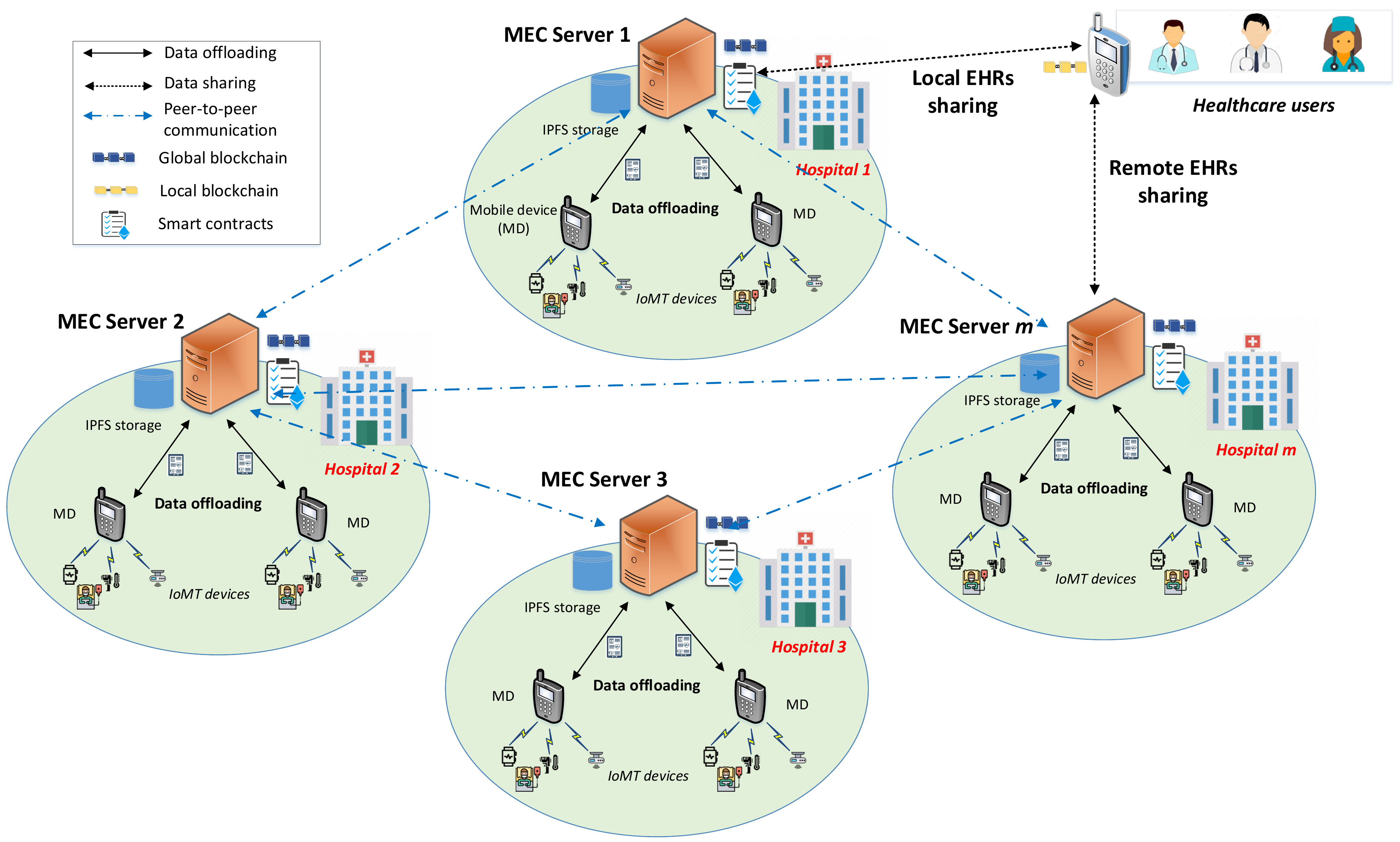} 
	\caption{The proposed healthcare architecture with MEC and blockchain. }
	\label{Fig:Overview}
	\vspace{-0.1in}
\end{figure*}
\subsection{Health Data Sharing}
Several solutions using blockchain have been introduced for health data sharing. Our preliminary work in \cite{DinhGlobe} proposed a decentralized health data sharing based on blockchain and MEC for federated hospitals. The work \cite{11} presented a privacy-preserved data sharing scheme enabled by the conjunction of a tamperproof consortium blockchain and cloud storage. Furthermore, the authors in \cite{12} described a hybrid architecture of using both blockchain and edge-cloud nodes where smart contracts are employed to monitor access behaviours and transactions. Despite data privacy enhancements, such solutions \cite{11}, \cite{12} mainly rely on central cloud servers for EHRs storage, which remains single-point failure bottlenecks and incurs high communication overhead. Further, the performances of smart health contract have not been evaluated. Another work in \cite{13} employed an IPFS system with Ethereum blockchain for EHRs sharing over clouds, but the important performance metrics such as data retrieval speed and security capability have not been verified. Moreover, the work in \cite{30} described a privacy-preserving EHRs sharing by using blockchain in the Ethereum. Blockchain stores hash values of EHRs while raw data is kept on the cloud server. The authors also constructed a stealth authorization framework to achieve privacy-preserving access authorization delivery over the hospital network. The study in \cite{31} suggested an EdgeCare model that uses edge computing for healthcare management in decentralized health environments. The local authorities were employed to perform access verification for the data offloading and sharing, but healthcare storage has not been considered. Also, a privacy-preserving sharing scheme for IoTs including healthcare was also considered in \cite{32} by using a classic IPFS system with a global DHT look-up solution that results in unnecessary
communication overhead.


\section{System Model}
\label{SystemModel}
In this section, we present a new health architecture in hospital networks and then describe our blockchain design. 
\subsection{Proposed Healthcare Architecture with MEC and Blockchain}
Here, we propose a decentralized health architecture as shown in Fig.~\ref{Fig:Overview}. The proposed architecture consists of a network of cooperative hospitals linked by a blockchain ledger. Each hospital is controlled by an MEC server which executes the data tasks offloaded by MDs and also communicates with other MEC servers for data sharing. Here, we consider a set of MEC servers as $\mathcal{M} = \{1,2,..., M\}$, each MEC server $m$ is located at a hospital $HP_m$. We assume that each hospital has a set of MDs $\mathcal{N} = \{1,2,..., N\}$ which collect data from a set of patient $\mathcal{J} = \{1,2,..., J\}$ using IoMT devices. Then, this data is offloaded to the MEC server for computation via the data offloading scheme. Note that here we consider at each hospital $HP_m$, a MD $n$ can collect health data from different patients. We also assume that there is a set of healthcare users (HUs) $\mathcal{U} = \{1,2,..., U\}$ such as doctors, clinicians who may be situated at any hospital and want to access data on the MEC network via the data sharing scheme (healthcare users, users and HUs are used interchangeably throughout the paper). The details of each network components are explained as follows.
\begin{itemize}
	\item \textbf{MEC servers:} Each MEC server acts as a coordinator that manages a group of MD devices in its hospital to provide low-latency computation services in the data offloading scheme. The MEC server also links with the other MEC servers from other hospitals in a peer-to-peer (P2P) manner to build a decentralized data sharing network. \textcolor{black}{We consider a realistic scenario that MEC servers may be semi-trusted, which means that the MEC server may be curious about health data and misbehaves in the data sharing. To overcome these challenges, we store sensitive EHRs data in the IPFS, instead of in the MEC server's hard drive, while the hash value of data record is kept in the smart contract so that any data retrieval behaviours are traced on the blockchain in a transparent manner.}
	\item \textcolor{black}{\textbf{Healthcare users (HUs):} HUs such as doctors and clinicians across federated hospitals may be interested in EHRs data for providing healthcare, e.g., medical diagnosis. Considering the realistic health settings that the HUs are not trusted, we develop an access control mechanism based on smart contracts to perform authentication of user access behaviours, which will be presented in Section~\ref{Sharing}. Only authenticated users are able to retrieve data in the IPFS, while malicious users are detected and prevented.}
	\item \textbf{Mobile devices (MDs):} MDs such as smartphones, laptops are responsible for gathering data from IoMT sensors. The MD can use its hardware to execute its data tasks or offload them to a nearby MEC server by data offloading. In this paper, we use smartphones as MDs to collect data from wearable sensors and perform data computation. 
	\item \textbf{IPFS storage:} {As MEC servers are semi-trusted, in our paper, raw EHRs collected from IoMT and data processed from the offloading phase} are uploaded to the decentralized off-chain IPFS platform running on top of the MEC network. IPFS introduces low-latency and fast decentralized archiving with reliable P2P content delivery, which has been investigated in healthcare scenarios \cite{25}; and thus it is well suitable for our health applications. Specially, in our work, we make an improvement in the IPFS design by replacing the global DHT with smart contracts to manage the hash records of health data, aiming to solve communication latency issues remained in classic IPFS systems \cite{13,14}.
	\item \textbf{Smart contract:} A smart contract can be regarded as a self-operating computer program which is automatically executed when specific conditions are met \cite{mcghin2019blockchain}. Each contract has an account that contains data and codes with programmable logic functions. In our paper, we design a contract called access control smart contract (ACSC). Each MEC server holds a copy of smart contracts and any their new events (i.e., user access) are updated at other MEC servers via the global blockchain network. The details of our contract design will be explained in the following section. 
\end{itemize}
\subsection{Blockchain Design}
Blockchain is the heart of our decentralized healthcare architecture. In this paper, we suggest using a permissioned Hyperledger Fabric \cite{26} blockchain platform based on practical Byzantine fault tolerance (PBFT) consensus to implement our healthcare system. The Hyperledger Fabric blockchain only allows authenticated users to join the network, and the validation is performed by only pre-selected nodes with high computing capability, i.e., MEC servers in our scenario, without mining requirements for lightweight entities like mobile users. This would improve the transaction performance, i.e., low transaction latency, compared to permissionless blockchains such as Ethereum \cite{26}. As shown in Fig.~\ref{Fig:Overview}, we consider two types of blockchain for our healthcare architecture: a global blockchain and local blockchains. 
\begin{itemize}
	\item \textbf{Global blockchain:} It interconnects all MEC servers together for hospital communications under the control of all MEC servers. Once a data storage event at each MEC server occurs (triggered by the offloading process), this MEC server broadcasts an offloading transaction to the other MEC servers for global updates. Moreover, in the data sharing phase, when a mobile user performs a data request to an MEC server, this MEC server also creates a sharing transaction and broadcasts it to other MEC servers in the hospital network.
	\item \textbf{Local blockchain:} Each hospital deploys a local hospital to link the local MEC server with its mobile users. This local blockchain is controlled by the local MEC server. When a mobile user performs a data request to the MEC server, he creates a sharing transaction and submits it to the local blockchain so that the MEC server can process the request and return the data. If the MEC server can look up data locally, the server will return immediately to the user. Otherwise, it asks the other MEC servers to find the address of the requested data and then responds the user. 
\end{itemize}
\textbf{Remark:} It is worth noting that in the proposed healthcare architecture as illustrated in Fig.~\ref{Fig:Overview}, blockchain is only used in the sharing scheme, while we assume that the offloading network is private and thus does not need to apply blockchain in the offloading scheme. In fact, this assumption holds in practical hospital settings \cite{banerjee2014monitoring} where each MD is managed by its healthcare provider such as a physician who uses the device to only collect data from his trusted patients and offload it to the MEC server in each medical test.

In the following sections, we will present our designs for a data offloading scheme and a data sharing scheme in details.

\section{Health Data Offloading}
\label{Offloading}
In this section, we introdude the offloading scheme and then formulate the offloading model in details. 
\subsection{Offloading Model}
\begin{figure}
	\centering
	\includegraphics[width=0.95\linewidth]{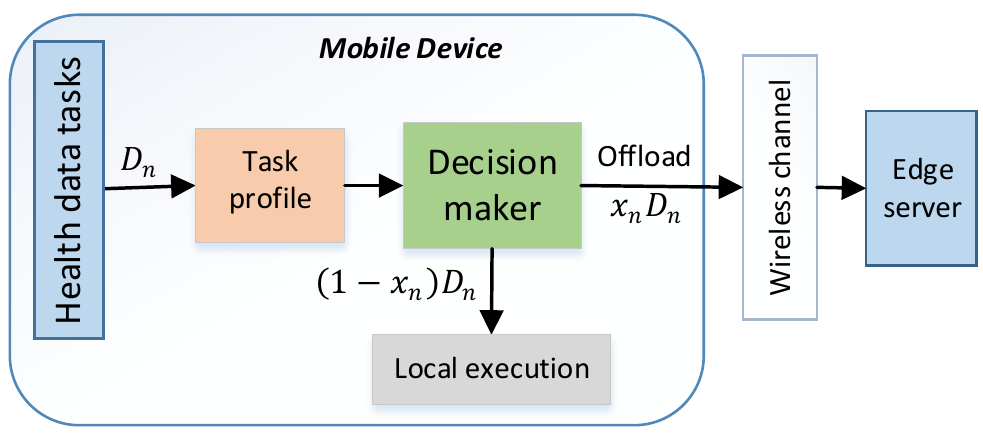}
	\caption{The data offloading scheme.  }
	\label{Fig:Offloading}
	\vspace{-0.1in}
\end{figure}

Here, we focus on formulating the offloading model for a representative hospital. Based on the QoS requirements (i.e., latency, energy consumption), each health data task can be executed at the MD or at the MEC server via task offloading.  Accordingly, we introduce an offloading decision policy for each MD $n$ denoted by a binary variable $x_n \in \{0,1\} $. Here $x_n= 0$ means that the MD $n$ decides to process its data task locally, and $x_n = 1$ indicates that the MD $n$ offloads its task to the MEC server. For the sake of simplicity, we assume that each MD $n$ has a task $Y_n$ to be executed, which can be defined by a variable $Y_n = (D_n, X^l_n, T^{max}_n)$, where $D_n$ expresses the size of the input data (in bit), $ X^l_n$ denotes the required number of CPU cycles of the task, and $ T^{max}_n$ specifies the maximum permissible latency (in second) to accomplish the task $ Y_n$. Motivated by the experimental results of our recent work \cite{16}, in this paper we propose an offloading architecture as shown in Fig.~\ref{Fig:Offloading} which consists of two main modules: task profile and decision maker installed on each MD $n$.

- \textit{Task profile}: This module is responsible for collecting all task information such as the data task $Y_n$, energy consumption ($E_n$), processing time ($T_n$) and memory usage ($M_n$), by using mobile performance measurement tools. Therefore, a task profile of a MD $n$ can be formulated as a tuple $ [Y_n, E_n, T_n, M_n]$ which is then stored in a database of the MD $n$ for supporting the offloading decision process. Details of the modelling of each component in the task profile will be presented in the following section. 

- \textit{Decision maker}: This module receives task profile information collected by the profile module to make offloading decisions. Similar to \cite{18}, we employ an integer linear programming model to develop a decision making algorithm on MDs. By using profile information, the algorithm analyses and makes decisions whether the data task should be offloaded to the MEC server or not. The main objective is to determine an optimal computation decision for each task to minimize the offloading cost of energy consumption, execution latency and memory usage. 

\subsection{	Offloading Formulation}
\textcolor{black}{In the literature, the existing schemes \cite{5,9,8,10} lack a joint consideration of important QoS metrics in health data offloading, including processing time, energy consumption and memory usage. Moreover, data privacy issues in the offloading have not been solved \cite{4}. Motivated by the limitations of current schemes, we here formulate the health data offloading problem by taking these factors into account.} Two computation modes are considered, namely local execution and offloading.
\subsubsection{Local Execution}
When a MD decides to execute the task $Y_n$ locally ($x_{n} =0$), it uses its resource to process healthcare data. We denote $X^l_n,f^l_n$ as mobile CPU utilization for task $n$ (in CPU/bit) and mobile computational capacity (in CPU/sec), respectively. Then, the local execution time can be calculated as 
\begin{equation}
T^{local}_n = \dfrac{D_nX^l_n}{f^l_n}.
\end{equation}

We also define $E^{local}_n$ and $M^{local}_n$ as battery consumption (in Mah) and memory usage (in MByte), which can be measured by mobile measurement tools \cite{18}. 

\subsubsection{Offloading to the MEC Server} In the case of task offloading ($x_{n} =1$), the data task needs to be encrypted for privacy before transmitting to the MEC server. \textcolor{black}{Here we employ an Advanced Encryption Standard (AES) encryption function installed in each MD to encrypt healthcare data due to its less energy consumption and low latency \cite{kumar2017comparative}.} Thus, the time required to accomplish a task in the offloading mode includes three parts: the task encryption time at the MD $n$, the time for transmitting the task to the MEC server, and the task execution time at the MEC server. Then, let denote $X^{enc}_n, X^e_n $ as mobile CPU utilization for encrypting the task $Y_n$ (in CPU cycles/bit) and edge CPU utilization (in CPU cycles/bit), respectively, the total offloading time is specified as
\begin{equation}
T^{offload}_n = \dfrac{D_nX^{enc}_n}{f^l_n} + \dfrac{D_n}{R_n} + \dfrac{D_nX^{e}_n}{f^e_n},
\end{equation}
where $f^e_n$ is defined as the computational resource (in CPU cycles/second) allocated by the MEC server to accomplish the task $Y_n$, which should not exceed the available edge computational budget $f^e$. Also, $R_n$ is the data transmission rate of the MD $n$ (in bit/second).

Moreover, the energy consumption for the offloading is mainly characterized by the encryption energy and transmission energy \cite{18}. Then, we define $E^{enc}_n$ as battery consumption when encrypting the task $Y_n$ at the MD $n$ which is measured by mobile measurement tools. Moreover, according to \cite{16}, the transmission energy is computed by
\begin{equation}
E^{trans}_n = p_n\dfrac{D_n}{R_n},
\end{equation}
where $p_n$ is the transmission power of the MD $n$. Hence, the total offloading energy is specified by
\begin{equation}
E^{offload}_n =E^{enc}_n + E^{trans}_n. 
\end{equation}
The offloading process also incurs a memory usage cost for encryption, defined as $M^{offload}_n$, which can be also obtained through mobile measurement tools \cite{18}. Accordingly, the total offloading time, energy consumption and memory usage can be expressed respectively as
\begin{equation}
T_n = (1-x_n)T^{local}_n + x_nT^{offload}_n,
\end{equation}
\begin{equation}
E_n = (1-x_n)E^{local}_n + x_nE^{offload}_n,
\end{equation}
\begin{equation}
M_n = (1-x_n)M^{local}_n + x_nM^{offload}_n.
\end{equation}
\subsubsection{Offloading Problem Formulation}
Based on above formulations, we derive the optimization problem to jointly optimize time latency, energy consumption and memory usage under system constraints as follows
\begin{equation}
\begin{aligned}
& (P1): \underset{\textbf{x}}{\min}
&& \sum_{n=1}^{N} \left(\alpha_tT_n+\alpha_eE_n+\alpha_mM_n\right)\\
& \text{subject to}
&& (C1): x_n \in \{0,1\}, \forall n \in \mathcal{N}, \\
&&& (C2): T_n \leq T^{max}_n, n \in \mathcal{N},\\
&&& (C3): R_n \geq R^{min}_n, n \in \mathcal{N},\\
&&& (C4): M_n \leq M^{max}_n, n \in \mathcal{N}, \\
&&&(C5): 0 < p_n \leq P_n, n \in \mathcal{N},  \\
&&& (C6): 0 <f^e_n \leq f^e, n \in \mathcal{N}.
\end{aligned}
\end{equation}
where $\textbf{x} = [x_1, x_2,...,x_N]$ is the offloading decision vector of all MDs $n \in \mathcal{N}$, and $\alpha_t, \alpha_e, \alpha_m$ are the weights of time, energy and memory cost, respectively. In the problem \textit{(P1)}, the constraint (C1) implies that each task can be either executed locally or offloaded to the MEC server.  Further, the total task execution time should not exceed a maximum latency value (C2). The constraint (C3) is added to guarantee the reliability of the task offloading at a MD $n$, where $R^{min}_n$ is the minimum transmission rate and specified by $ R^{min}_n = D_n/T^{max}_n$. This condition ensures that the transmission time of MD $n$ during the offloading process is not too long, and thus the computation of the data task can be completed before the deadline $ T^{max}_n$. (C4) defines that the memory used for task computation must not exceed the available mobile memory. (C5) shows the transmission power constraint of each MD, and the constraint (C6) shows that the MEC server must allocate a positive computing resource to each MD under an available computation budget $f^e$. In this paper, we employed a particle swarm optimization (PSO) \cite{19} model written in java to build the above offloading optimization algorithm in Android phones. The PSO algorithm has proven its superior advantages over its counterparts like genetic algorithm in terms of extremely low computational cost and simple implementation on Android devices for mobile offloading applications like healthcare \cite{elkady2016modified}. 

\subsection{Health Data Storage}
The raw health data associated with the analysed results from the data offloading scheme are stored in the IPFS system running on top of the MEC network. We here analyze a representative hospital $HP_m$ where there is an MEC server $m$ along with its network of MDs $n$ ($n \in \mathcal{N}$) and patients $PID_j$ ($j \in \mathcal{J}$). The data storage process is implemented through four steps.
\begin{enumerate}
	\item The raw health data of the patient $PID_j$ collected from the MD $n$ is offloaded to the MEC sever $m$ for computation. {For simplicity, here we assume that each patient $j$ has a health data record.} Then, the MEC server aggregates the raw data and the analysed result and adds them into a data record identified by the patient ID $j$ as
	\begin{equation}
	C_j = (rawdata||analyzedResult).
	\end{equation} 
	\item Instead of storing the data and analysed results in the MEC server's hard drive, we encrypt these data as
	\begin{equation}
	C^{enc}_j \leftarrow Enc(C_j,PKM_m),
	\end{equation}
	where $PKM_m$ is the private key of the MEC $m$ (its creation will be explained in the following section). Then we put this data to the IPFS storage node located on top of the MEC server for data security.
	\item Uploading the data to the IPFS storage would automatically return a cryptographic hash of its content by using a hash function $H_{IPFS}$:
	\begin{equation}
	h_j = H_{IPFS}(C^{enc}_j, timestamp).
	\end{equation}
	Here, we keep the hash value in the smart contract on the MEC server for fast data look-up, instead of relying on the classic DHT.
	\item \textcolor{black}{The storage of a data record in the IPFS node at the MEC server $m$ is synchronized with other IPFS nodes at other hospitals via the P2P network for global data sharing. The MEC server $m$ also adds ($h_j, PID_j, PKM_m, timestamp$) as a transaction including the EHRs' hash $h_j$ and broadcasts it to the global blockchain network: 
		\begin{equation}
		MEC_m \rightarrow *: (h_j, PID_j, PKM_m, timestamp).
		\end{equation} 
		The other MEC servers will receive the transaction and extract the offloading information $h_j, PID_j$ that is then stored on the database of the ACSC contract.} The details of contract design will be presented in the following section.
\end{enumerate}

\section{Health Data Sharing}
\label{Sharing}
\textcolor{black}{The health data records stored in the IPFS from the data offloading scheme are then used in the data sharing scheme in federated hospitals. To provide authentication for data sharing, a smart contract-based access control solution is adopted at the network edge without requiring any central authority like previous works \cite{11}, \cite{12}. This not only reduces authentication latency but also increases the reliability of data sharing. Moreover, the integration of smart contract into IPFS enables the replica of hash value of data records across the edge network which helps accelerate the data retrieval rate while the data traceability is achieved.} In the following, we explain our proposed data sharing scheme as shown in Fig.~\ref{Fig:Overview}, including two key parts: user authentication and data retrieval.
\subsection{User Authentication with MEC and Smart Contract}
The user authentication process consists of three phases: initialization phase, user registration phase, and data retrieval phase, which are explained as follows.
\subsubsection{	Initialization Phase}
Each hospital is initilized by its MEC server. In this phase, a key is set up by the MEC server for data sharing establishment. Also, the MEC server $m$ in each hospital sets up the private key and the public key. 
\subsubsection{Registration Phase}
This phase is invoked whenever a HU wants to register to the MEC server for the first time. To do so, the user joins the local blockchain network and follows the steps as below.
\begin{itemize}
	\item Each HU $u$ submits a transaction $T_{{reg}_u}$ to the MEC server $m$ as $T_{{reg}_u}$ for registration:
	\begin{equation}
	T_{{reg}_u} = (PKU_u||ID_u||timestamp).
	\end{equation} 
	\item The MEC server $m$ decodes the $T_{{reg}_u}$ to obtain the user's public key as $PKU_u \leftarrow T_{{reg}_u}.getSenderPublicKey()$.
	\item  Then, the MEC server decodes the transaction to get the user ID. First, it decodes the transaction $T_{{reg}_u}$, then finally obtain the user ID as an unique address $ADD_u$:
	
	\begin{equation}
	deT_u \leftarrow abiDecoder.decodeMethod(T_{{reg}_u}),
	\end{equation}
	\begin{equation}
	ADD_u \leftarrow  web3.eth.getData(deT_u([ID_u])).
	\end{equation}
	\item The MEC server checks user information and stores $\{ PKU_u, ADD_u \}$ as the user identification information in the contract database.
	\item The MEC server then calculates the hash value of the register transaction $T_{{reg}_u}$ as:
	\begin{equation}
	H_{{HU}_u} \leftarrow Hash(T_{{reg}_u}, SKM_m),
	\end{equation}
	which is then published to the local hospital blockchain network for tracing. The MEC server $m$ also broadcasts its public key $PKM_m$ to the user that is necessary for the future user data access.
\end{itemize}
\subsubsection{User Authentication Phase}
It is supposed that a HU $u$ wants to access the patient's EHRs stored on the MEC network for their medical tasks. To obtain the EHRs of the target patient, the HU $u$ needs to know his patient identity $PID_j$ so that the MEC server can locate the address of this patient in the hospital network during EHRs sharing. The data retrival process is presented by the following key steps.
\begin{itemize}
	\item A HU $u$ prepares a data retrieval request $T_{{req}_u}$ involved a target patient ID $PID_j$ and the address of patient's hospital $HP_w$ ($w \in \mathcal{M}$). Thus, the target patient address on the hospital network can be expressed as $P_{addr}= <PID_j,HP_w>$, i.e., the $5^{th}$ patient in the $3^{th}$ hospital.  Then, the request $T_{{req}_u}$ can be specified by 
	\begin{equation}
	T_{{req}_u} \leftarrow (PKU_u||ID_u||P_{addr}||timestamp),
	\end{equation}
	where each compoment in $T_{{req}_u}$ is formatted with an index in the array $index=[1-4]$, i.e., the index of $PKU_u$ is 1. This format is necessary for transaction decoding later. 
	\item To ensure privacy, the user request should be encrypted with the MEC $m$'s public key $PKM_m$ (obtained from the registratrion phase) as $T_{{enc}_u} \leftarrow Enc(T_{{req}_u},PKM_m)$ and submits it to the MEC $m$.
	
	\item At the edge side, the MEC server $m$ decrypts the user request $T_{{enp}_u}$ as $T_{{dec}_u} \leftarrow Dec(T_{{req}_u},SKM_m)$. 
	To provide security for the EHRs sharing, user authentication is highly essential. To do so, the MEC server extracts the user's public key from the request as
	\begin{equation}
	Pub_u \leftarrow T_{{dec}_u}.getSenderPublicKey().
	\end{equation}
	It also decodes the transaction $T_{{dec}_u}$ to obtain the request information $ReqInf_u$: 
	\begin{equation}
	deT_{{dec}_u} \leftarrow abiDecoder.decodeMethod(T_{{dec}_u}),
	\end{equation}
	\begin{equation}
	ReqInf_u \leftarrow  web3.eth.getData(deT_{{dec}_u}[DataIndex]),
	\end{equation}
	and then obtains the user identity $Iden_u$ as
	\begin{equation}
	Iden_u = ReqInf_u(Index[index_{Iden}]).
	\end{equation}
	\item The MEC server will check and authenticate the received user identification information $<Pub_u,Iden_u>$, and then put them into user mapping as
	\begin{equation}
	UMAP_{PK_{u}} = Map<Pub_u => PKU_u>,
	\end{equation}
	\begin{equation}
	UMAP_{ID_{u}} = Map<Iden_u => ID_u>,
	\end{equation}
	by using the smart contract (see in Algorithm 1). If both $UMAP_{PK_{u}} \rightarrow true$ and $UMAP_{ID_{u}} \rightarrow true$, the user request is validated successfully, otherwise a penalty is issued for
	access prevention.
	
	\item In the case of successful request validation, the MEC server $m$ will calculate the signature of $T_{{dec}_u}$ as
	\begin{equation}
	Sig_u \leftarrow Hash(T_{{dec}_u}, SKM_m),
	\end{equation}
	Finally, the MEC server will issue a certificate $Cert_{u}$ as 
	\begin{equation}
	Cert_{u}= \{ Sig_u, PKU_u, ID_{u}, timestamp\},
	\end{equation}
	which is then sent to the HU $u$ via the local blockchain for successful authentication proof. 
\end{itemize} 

\subsection{Health Data Retrieval with MEC and Blockchain}
After successful authentication, the MEC server $m$ will locate the requested EHRs based on the patient information $<PID_j,HP_w> \leftarrow ReqInf_u(Index[indexPID_u])$ that is defined in the authentication phase. In fact, the patient and the user may be located in the same hospital or in different hospitals. For example, a patient may only stay at a hospital for his treatment, but in some cases (e.g., seeking treatment for a new disease), he wants to visit a different doctor in another hospital. Motivated by this realistic scenario, here we consider two cases: (1) The patient and the user are in the same hospital and (2) The patient and the user are in different hospitals.

\textbf{Case 1: The patient and the user are in the same hospital:} In this case, the MEC server finds that the HU and the patient are in the same hospital by checking $HP_w$ information. It is supposed that the HU $u$ communicates with the MEC server $m$ to request the data record of the patient $PID_j$ in the same hospital, then the data retrieval process is implemented by the following steps:
\begin{itemize}
	\item The MEC server $m$ first verifies the request information $PID_j$ by refering to the ACSC contract to perform mapping between the patient record stored in the contract $PID_j^{sc}$ and information in the request $PID_j^{req}$:
	\begin{equation}
	UMAP_{PID_{j}} = Map<PID_j^{sc} => PID_j^{req}>. 
	\end{equation}
	If $UMAP_{PID_{j}} \rightarrow true$, the request information is verified for ready data retrieval. 
	\item Based on the received patient information, the ACSC contract will extract the hash value $h_j$ that represents the patient $j$'s health record. Then, the contract sends a request to the IPFS storage for data retrieval using the hash by a command: $C_j^{enc} = GET_{IPFS}(h_j)$, i.e.,: {\fontfamily{pcr}\selectfont ipfs get /ipfs/ Qmd84db7be0690ebb015f1cD9d9491cE18076c}. 
	\item Since the data stored on the IPFS was encrypted in the storage process (see in Section IV-C), we need to decrypt to obtain the real data as
	\begin{equation}
	C_j \leftarrow Dec(C_j^{enc}, SKM_m).
	\end{equation}
	The MEC $m$ then returns the data $C_j$ via a secure channel to the requestor. 
	\item Finally, the MEC server $m$ adds a conjunction of ($PKU_u, h_j, PKM_m, timestamp$) as a transaction and broadcasts it to the global blockchain network: 
	\begin{equation}
	MEC_m \rightarrow *: (PKU_u, h_j, PKM_m, timestamp).
	\end{equation}
\end{itemize}
\begin{figure*}
	\centering
	\includegraphics [width=0.9\linewidth]{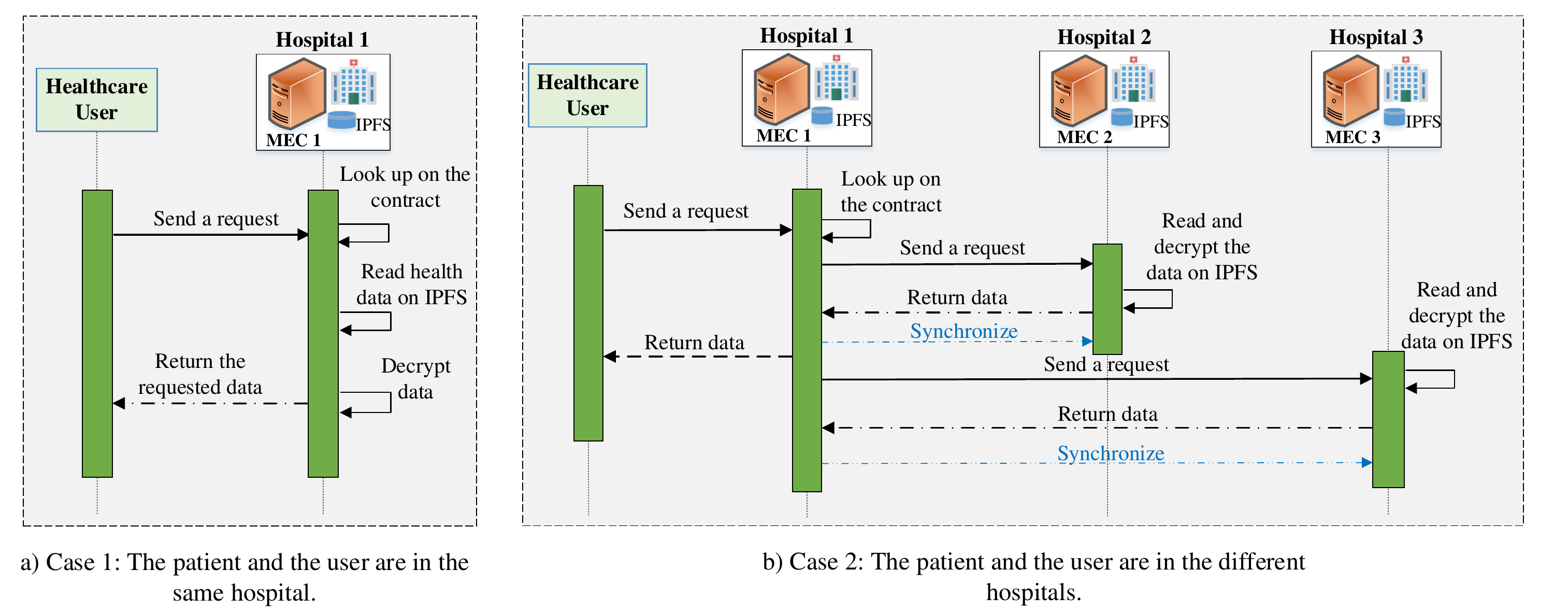}
	\caption{The proposed health data sharing procedure with MEC and IPFS. }
	\label{Fig:Sharing}
	\vspace{-0.15in}
\end{figure*}

\textbf{Case 2: The patient and the user are in different hospitals:} 
In this case, the MEC server will seek the address of the patient in the MEC network. \textcolor{black}{Due to all patient addresses ($PID_j,HP_w$) are stored on the ACSC contract replicated across the global hospital network, an MEC server at a hospital can know exactly where the requested patient data is currently located. Hence, an MEC server only needs to send the data request to the destination MEC server (using $HP_w$ information) that contains the requested data for data retrieval, without broadcasting the request to all MEC servers.} This strategy not only saves data lookup time but also saves network bandwidth and potentially reduces the traffic congestion on the global blockchain network. The data retrieval process in this case is summarized as the following steps:
\begin{itemize}
	\item The MEC server $m$ first also verifies the request information $PID_j$ from the user $PKU_u$ by refering to the ACSC contract, and then performs mapping to verify that the patient information $PID_{j}$ is correct. Then, it also identifies which MEC server is currently storing the requested data by checking $HP_w$ information. Here, we assume that an MEC server $MEC_y, (y\#m, y \in \mathcal{M})$ is holding the requested data.
	\item After identifying the destination MEC server $y$, the MEC server $m$ will send a transaction for data retrieval request: 
	\begin{equation}
	MEC_m \rightarrow MEC_y: (PID_j, PKU_u, PKM_m, time).
	\end{equation}
	\item Based on the patient information $PID_j$, the ACSC contract in the MEC $y$ obtains the hash value $h_j$. Next, the contract sends a request to the IPFS node in the MEC $y$ by a command: $C_j^{enc} = GET_{IPFS}(h_j)$, which is then decrypted to obtain the real data:
	\begin{equation}
	C_j \leftarrow Dec(C_j^{enc}, SKM_y).
	\end{equation}
	\item The MEC server $y$ then transmits the collected data $C_j$ to the MEC server $m$ so that the server $m$ returns it to the requestor. Finally, the MEC server $y$ adds a conjunction of ($PKU_u, h_j, PKM_y, timestamp$) as a transaction and broadcasts it to the global blockchain network:
	\begin{equation}
	MEC_y \rightarrow *: (PKU_u, h_j, PKM_y, timestamp).
	\end{equation}
	Finally, all MEC servers update the user access events and acheive a synchronization over the data sharing across the hospital network.  
\end{itemize}
The data retrieval process for two cases are shown in Fig.~\ref{Fig:Sharing}, and the data sharing is summarized in Algorithm 1. Here, lines (5-8) present the pre-processing steps when a HU $u$ submits a request to the MEC server. Then, the MEC server will authenticate the request using the ACSC, by verifying the identification information ($PKU_u, ID_u$) (lines 9-18). For the request authenticated by the contract, the MEC server specifies whether the patient and the user are in the same hospital or in the different hospitals. The MEC server will refer to the IPFS storage for data retrieval based on patient information $PID_j$ before returning to the HU $u$ (in lines 19-38). 
\begin{algorithm}
	\footnotesize
	\caption{Health data sharing with MEC and blockchain}
	\begin{algorithmic}[1]
		\STATE \textbf{Input:} $ PID_j, HP_w, PKU_{u}, ID_u, T_{{req}_u} $
		\STATE \textbf{Output:} $ Auth_u, C_j$
		\STATE \textbf{Initialization:} \textit{(by the user HU $u$)} 
		\STATE Encrypt the request: $T_{{enc}_u}$ and submits it to the MEC $m$
		\STATE \textbf{Pre-processing the request} \textit{(by MEC)} 
		\STATE The MEC $m$ decrypts the user request $T_{{enp}_u}$ as $T_{{dec}_u} \leftarrow Dec(T_{{req}_u},SKM_m)$
		\STATE Obtain the user's public key: $Pub_u \leftarrow T_{{dec}_u}.getSenderPublicKey()$
		\STATE Decode $T_{{dec}_u}$ and get the user identity $Iden_u$
		\STATE \textbf{Authentication} \textit{(by the ACSC contract)} 
		\IF {$msg.sender == MEC_m$}
		\STATE $PKcheck = policy[EHRresource][action].PKU_u$
		\STATE $Idencheck = policy[EHRresource][action].ID_u$
		\IF {$PKcheckIdencheck \rightarrow true$}
		\STATE $Auth_u \leftarrow AccessResult(msg.sender, Accepted, true, time)$ \\ Accept the user request
		\ELSE 
		\STATE $Auth_u \leftarrow AccessResult(msg.sender, Denied, false, time)$ \\ Refuse the user request
		\ENDIF
		\ENDIF
		\WHILE {$Auth_u \rightarrow true$} 
		\IF {$HP_w == HP_w^{sc}$} 
		\STATE (\textit{{\textbf{Case 1}: The patient and the user are in the same hospital}})
		\IF {$PID_j^{sc}==PID_j$}
		\STATE Get the data on IPFS: $C_j^{enc} = GET_{IPFS}(h_j)$
		\STATE Decrypt to obtain the real data $C_j \leftarrow Dec(C_j^{enc}, SKM_m)$
		\ENDIF
		\STATE The MEC $m$ returns the data $C_j$ to the $HU_u$
		\STATE The MEC $m$ adds a transaction to the global blockchain network: $MEC_m \rightarrow *: (PKU_u, h_j, PKM_m, timestamp)$
		\ELSIF {$HP_w != HP_w^{sc}$} 
		\STATE (\textit{{\textbf{Case 2}: The patient and the user are in different hospitals}})
		\IF {$PID_j^{sc}==PID_j$}
		\STATE Communicate with the MEC $y$: $MEC_m \rightarrow MEC_y: (PID_j, PKU_u, PKM_m, timestamp)$
		\STATE Get the data on IPFS: $C_j^{enc} = GET_{IPFS}(h_j)$
		\STATE Decrypt to obtain the real data $C_j \leftarrow Dec(C_j^{enc}, SKM_y)$
		\ENDIF
		\STATE The MEC $y$ returns data $C_j$ to the MEC $m$ and then the HU $u$
		\STATE The MEC $y$ adds a transaction to the global blockchain network: $MEC_y \rightarrow *: (PKU_u, h_j, PKM_y, timestamp)$
		\ENDIF
		\ENDWHILE
	\end{algorithmic}
	
\end{algorithm}
\vspace{-0.1in}

\section{Experimental Results and Evaluations}
\label{Experimental}
In this section, we present experiments and perform implementation evaluations in details. 
\subsection{Experiment Settings}

We implemented a testbed to evaluate the proposed \textit{BEdgeHealth} architecture. We employed three computers (Microsoft Windows 10 64-bit, Intel core i7 at 3.4 GHz, 8GB of RAM) to work as MEC servers where each MEC server represents a hospital. Each server will connect with a network of MDs via a Cisco access point. Here we used Sony Android mobile phones as MDs for data offloading and data sharing tasks with Qualcomm Snapdragon 845 processor, 2GB memory, and a battery capacity of 2870mAh. In our study, healthcare data and programming code are necessary. For a specific case study, we used Biokin sensors \cite{14} developed by our lab as IoMT devices to collect simultaneously human motion data (including acceleration and gyroscope time-series data) with the size from 100 KB to 5 MB. This data is stored in separate files to be executed by an analysis program integrated in both Android phones and the MEC server. By using a programmed data analysis algorithm, we can estimate the movement disorder of a patient to serve doctors during clinical decisions \cite{14}. The experiment parameters are set up as follows: task CPU workload $ X_n^l = [0.8-1.5]$ Gcyles, maximum latency threshold $T_n^{max}=10$ second, device computing capability $f^l_n= 1$ GHz, edge computing capability $f^e = 5$ GHz. The maximum transmission power $P_n$ of MDs is set to $20$ mW, and the minimum transmission rate requirement $R^{min}_n = 0.5$ Mbps. For mobile performance evaluations, we employed a Firebase Performance Monitoring\footnote{https://firebase.google.com/docs/perf-mon} service to measure processing time, battery consumption, and memory usage. The mobile application for the offloading optimization was implemented using Android studio 3.5. 

For blockchain deployment, two Hyperledger Fabric platforms version 1.3 were used to build the global blockchain on the MEC system and the local blockchain on the MEC-device system. The PBFT consensus was implemented by MEC servers, while devices only joined the blockchain network for data request. We followed the instructions in the official Hyperledger Fabric tutorial to install required files and docker images \cite{27}. The smart contract was implemented in docker to serve user authentication and data retrieval \cite{28}. We also installed the JavaScript version of the IPFS platform in the MEC system \cite{29}.  Each of three MEC servers holds an IPFS node which is embedded with the Fabric blockchain to perform data storage and data sharing. To highlight the merits of the proposed \textit{BEdgeHealth} architecture, we compare our scheme with the related works using different performance metrics which are presented as follows. 
\subsection{Evaluation of Data Offloading Performance}
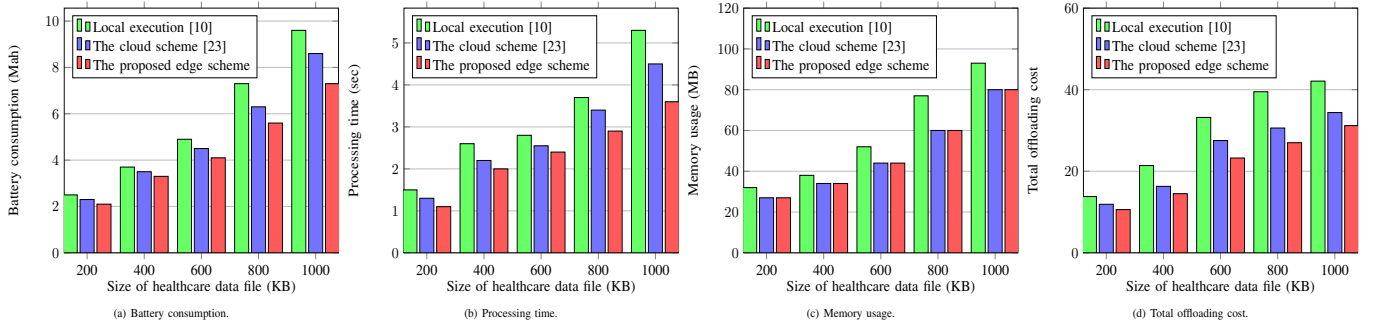
\begin{figure*}
	\centering
	\begin{adjustbox}{ height=4.1cm, width= 18cm}
		\subfigure[Battery consumption.]{\begin{tikzpicture}
			\begin{axis}[ybar,legend pos=north west, legend cell align={left}, 
			ylabel=Battery consumption (Mah),ymajorgrids,
			xlabel = Size of healthcare data file (KB),
			symbolic x coords={200, 400, 600, 800, 1000},
			xtick=data,ymin=0
			]
			\addplot [fill=green!60] coordinates  {(200,2.5)  (400,3.7)  (600,4.9)  (800,7.3)  (1000,9.6) };
			
			\addplot [fill=blue!55] coordinates  {(200,2.3)  (400,3.5)  (600,4.5)  (800,6.3)  (1000,8.6) };
			\addplot [fill=red!65] coordinates  {(200,2.1)  (400,3.3)  (600,4.1)  (800,5.6)  (1000,7.3)  };
			
			\legend{Local execution \cite{14}, The cloud scheme \cite{18}, The proposed edge scheme}
			\end{axis}
			\end{tikzpicture}}
		\subfigure[Processing time.]{\begin{tikzpicture}
			\begin{axis}[ybar,legend pos=north west, legend cell align={left},
			ylabel=Processing time (sec),ymajorgrids,
			xlabel = Size of healthcare data file (KB),
			symbolic x coords={200, 400, 600, 800, 1000},
			xtick=data,ymin=0
			]
			\addplot [fill=green!60] coordinates {(200,1.5) (400,2.6)   (600,2.8) (800,3.7) (1000,5.3)   };
			
			\addplot [fill=blue!55] coordinates {(200,1.3) (400,2.2)   (600,2.55) (800,3.4) (1000,4.5)   };
			
			\addplot [fill=red!65] coordinates {(200,1.1)  (400,2)  (600,2.4)  (800,2.9)  (1000,3.6) };
			
			\legend{Local execution \cite{14}, The cloud scheme \cite{18}, The proposed edge scheme }
			\end{axis}
			\end{tikzpicture}}	
		\subfigure[Memory usage.]{\begin{tikzpicture}
			\begin{axis}[ybar,legend pos=north west, legend cell align={left},
			ylabel=Memory usage (MB),ymajorgrids, ymax=120,
			xlabel = Size of healthcare data file (KB),
			symbolic x coords={200, 400, 600, 800, 1000},
			xtick=data,ymin=0
			]
			\addplot [fill=green!60] coordinates  {(200, 32)  (400,38)  (600,52)  (800,77)  (1000,93)  };
			
			\addplot [fill=blue!55] coordinates  {(200,27)  (400,34)  (600,44)  (800,60)  (1000,80)  };
			
			\addplot [fill=red!65] coordinates  {(200,27)  (400,34)  (600,44)  (800,60)  (1000,80)  };
			
			\legend{Local execution \cite{14}, The cloud scheme \cite{18}, The proposed edge scheme}
			\end{axis}
			\end{tikzpicture}}
		\subfigure[Total offloading cost.]{\begin{tikzpicture}
			\begin{axis}[ybar,legend pos=north west, legend cell align={left},
			ylabel=Total offloading cost,ymajorgrids, ymax=60,
			xlabel = Size of healthcare data file (KB),
			symbolic x coords={200, 400, 600, 800, 1000},
			xtick=data,ymin=0
			]
			\addplot [fill=green!60] coordinates  {(200, 13.8)  (400,21.4)  (600,33.2)  (800,39.5)  (1000,42.1)  };
			
			\addplot [fill=blue!55] coordinates  {(200,11.9)  (400,16.3)  (600,27.53)  (800,30.6)  (1000,34.4)  };
			
			\addplot [fill=red!65] coordinates  {(200,10.6)  (400,14.5)  (600,23.25)  (800,27)  (1000,31.2)  };
			
			\legend{Local execution \cite{14}, The cloud scheme \cite{18}, The proposed edge scheme}
			\end{axis}
			\end{tikzpicture}}
	\end{adjustbox}
	\caption{\textcolor{black}{Comparison results for health data offloading with $\alpha_e =1/3, \alpha_t=1/3, \alpha_m=1/3$.}}
	\label{Fig:DataOffloadingResult1}
	\vspace{-0.15in}
\end{figure*}

\begin{figure*}
	\centering
	\begin{adjustbox}{ height=4.1cm, width= 18cm}
		\subfigure[Battery consumption.]{\begin{tikzpicture}
			\begin{axis}[ybar,legend pos=north west, legend cell align={left},
			ylabel=Battery consumption (Mah),ymajorgrids,
			xlabel = Size of healthcare data file (KB),
			symbolic x coords={200, 400, 600, 800, 1000},
			xtick=data,ymin=0
			]
			\addplot [fill=green!60] coordinates  {(200,2.2)  (400,3.4)  (600,4.4)  (800,7.14)  (1000,9.4) };
			
			\addplot [fill=blue!55] coordinates  {(200,1.95)  (400,2.8)  (600,4.2)  (800,6.04)  (1000,8.3) };
			\addplot [fill=red!65] coordinates  {(200,1.6)  (400,2.5)  (600,3.6)  (800,5.2)  (1000,6.7)  };
			
			\legend{Local execution \cite{14}, The cloud scheme \cite{18}, The proposed edge scheme}
			\end{axis}
			\end{tikzpicture}}
		\subfigure[Processing time.]{ \begin{tikzpicture}
			\begin{axis}[ybar,legend pos=north west, legend cell align={left},
			ylabel=Processing time (sec),ymajorgrids, ymax=11,
			xlabel = Size of healthcare data file (KB),
			symbolic x coords={200, 400, 600, 800, 1000},
			xtick=data,ymin=0
			]
			\addplot [fill=green!60] coordinates {(200,3.1) (400,5.3)   (600,6.2) (800,7.3) (1000,9.3)   };
			
			\addplot [fill=blue!55] coordinates {(200,2.4) (400,4.2)   (600,5.15) (800,6.9) (1000,8.2)   };
			
			\addplot [fill=red!65] coordinates {(200,1.9)  (400,3.5)  (600,4.4)  (800,5.5)  (1000,6.3) };
			
			\legend{Local execution \cite{14}, The cloud scheme \cite{18}, The proposed edge scheme }
			\end{axis}
			\end{tikzpicture}}
		\subfigure[Memory usage.]{\begin{tikzpicture}
			\begin{axis}[ybar,legend pos=north west, legend cell align={left},
			ylabel=Memory usage (MB),ymajorgrids,
			xlabel = Size of healthcare data file (KB),
			symbolic x coords={200, 400, 600, 800, 1000},
			xtick=data,ymin=0
			]
			\addplot [fill=green!60] coordinates  {(200, 62)  (400,75)  (600,105)  (800,135)  (1000,183)  };
			
			\addplot [fill=blue!55] coordinates  {(200,51)  (400,68)  (600,84)  (800,114)  (1000,156)  };
			
			\addplot [fill=red!65] coordinates  {(200,51)  (400,68)  (600,84)  (800,114)  (1000,156)  };
			
			\legend{Local execution \cite{14}, The cloud scheme \cite{18}, The proposed edge scheme}
			\end{axis}
			\end{tikzpicture}}
		\subfigure[Total offloading cost.]{\begin{tikzpicture}
			\begin{axis}[ybar,legend pos=north west, legend cell align={left},
			ylabel=Total offloading cost,ymajorgrids, ymax=60,
			xlabel = Size of healthcare data file (KB),
			symbolic x coords={200, 400, 600, 800, 1000},
			xtick=data,ymin=0
			]
			\addplot [fill=green!60] coordinates  {(200, 12.3)  (400,19.4)  (600,32.2)  (800,38)  (1000,38.3)  };
			
			\addplot [fill=blue!55] coordinates  {(200,11.2)  (400,15.3)  (600,19)  (800,24.4)  (1000,34.12)  };
			
			\addplot [fill=red!65] coordinates  {(200,10.6)  (400,12.5)  (600,16.6)  (800,20.25)  (1000,29.5)  };
			
			\legend{Local execution \cite{14}, The cloud scheme \cite{18}, The proposed edge scheme}
			\end{axis}
			\end{tikzpicture}}
	\end{adjustbox}
	\caption{\textcolor{black}{Comparison results for health data offloading with $\alpha_e =2/3, \alpha_t=1/6, \alpha_m=1/6$.}} 
	\label{Fig:DataOffloadingResult2}
	\vspace{-0.15in}
\end{figure*}
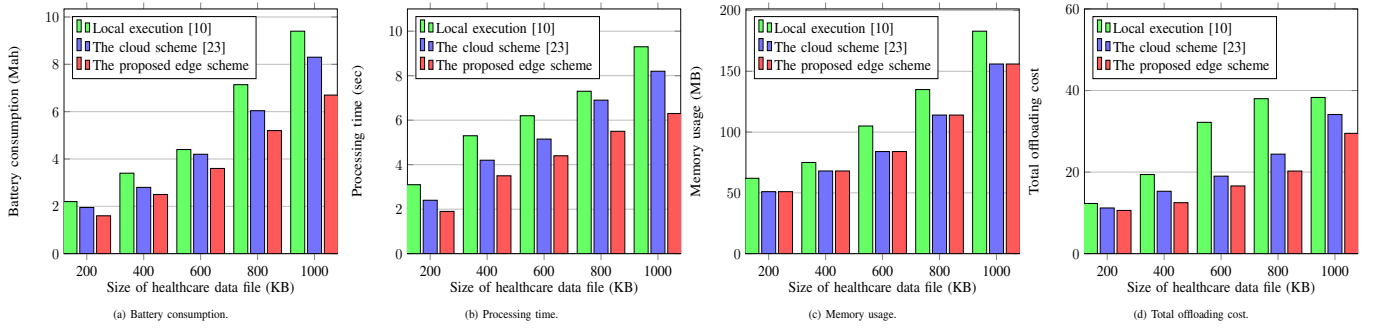

We compare our edge offloading scheme with two baselines: local execution \cite{14} (only executing data on devices) and cloud computation \cite{18} (offloading to the cloud server) to prove the advantages of our scheme. A set of health data files with different sizes (200~KB-1000~KB) collected from sensors is used in our evaluations. We implement each test with 10 times to obtain average values, and evaluate the offloading performance of these schemes on processing time, energy consumption, and memory usage. We present two settings with different weight values to evaluate the tradeoff among these three performance metrics, as shown in Fig.~\ref{Fig:DataOffloadingResult1} and Fig.~\ref{Fig:DataOffloadingResult2}. 

Fig.~\ref{Fig:DataOffloadingResult1} shows the offloading performances with $\alpha_e =1/3, \alpha_t=1/3, \alpha_m=1/3$. 
For the processing time, it consists of execution time in the local case, and encryption time, offloading time and remote execution time in the edge and cloud offloading cases. From Fig.~\ref{Fig:DataOffloadingResult1}(a), we can see that our edge offloading scheme consumes less energy when processing the health data tasks. As an example, offloading a 200~KB file  consumes in the edge scheme less 11\% energy than that in the case of local computation and less 5\% energy than that in the cloud scheme. Specially, the energy usage of the edge scheme becomes more efficient when the data size increases. For instance, the edge offloading scheme can save 21.3\% energy when executing a 1000~KB file, compared to the local scheme.

Fig.~\ref{Fig:DataOffloadingResult1}(b) indicates the average processing time of three schemes, with a significant performance improvement in the proposed edge scheme. For example, executing a 200KB file by the edge scheme only consumes 1.1 sec, whereas it requires about 1.3 sec and 1.5 sec in the cloud and local schemes, respectively. This leads to a 10-18\% time saving of data execution by using our edge scheme. Further, the proposed edge scheme saves up to 31\% and 15\% time when computing a 1000~KB file, compared to the local and cloud schemes, respectively. We also found with the selected human motion dataset, although data encryption is integrated in offloading, the edge-cloud offloading schemes still achieves better offloading performances than the local scheme, showing the efficiency of the proposed encryption technique.

Moreover, the memory performance is illustrated in  Fig.~\ref{Fig:DataOffloadingResult1}(c). The edge and cloud schemes have the same memory usage due to using the same encryption function for the offloading. However, these schemes achieve greater memory performances, with 5\% and 9\% memory savings compared to the local scheme when executing a 200~KB and 1000~KB file, respectively. Note that the above implementation results were obtained from the proposed offloading application with human motion data and current hardware settings of devices and MEC servers. Different mobile applications with other health data types such as videos and different hardware settings can achieve different offloading performances. However, in general, the proposed edge offloading scheme yields the best performances with reduced time latency, energy consumption, and better memory savings among three schemes for any task size. As a result, our edge scheme shows an significant improvement in the offloading performance with lower offloading cost, compared to the cloud scheme and the local scheme, as indicated in Fig.~\ref{Fig:DataOffloadingResult1}(d). 

Next, we investigate the offloading performances of three schemes with $\alpha_e =2/3, \alpha_t=1/6, \alpha_m=1/6$, as illustrated in Fig.~\ref{Fig:DataOffloadingResult2}. Based on the objective function in the optimization problem (\textit{P1}), a larger $\alpha_e$ value will give more penalty on the energy consumption. Particularly, the performance gap between the edge scheme and other schemes in this setting is larger than that in the previous setting in Fig.~\ref{Fig:DataOffloadingResult1} due to the increased latency and memory cost and reduced energy consumption. Moreover, our scheme achieves the best performance with lowest offloading cost among three schemes. For instance, when the task size is 1000~KB, the offloading cost of our edge scheme is 29.5, compared to 34.2 in the cloud scheme and 38.3 in the local scheme. The experimental results from two tradeoff settings in Fig.~\ref{Fig:DataOffloadingResult1} and Fig.~\ref{Fig:DataOffloadingResult2} demonstrate the advantage of our proposed edge scheme over the baselines, showing its efficiency in health offloading applications. 

\begin{table}
	\scriptsize
	\centering
	\captionsetup{font=scriptsize}
	\caption{Authentication cost test.}
	\begin{tabular}{|c||c|}
		\hline
		\textbf{Authentication functions}  & \textbf{Computation Cost } \\
		\hline
		Hash function 	&5.6 ms 
		\\
		Request encryption&	12.3 ms
		\\
		Transaction decoding&	3.2 ms 
		\\
		Request decryption&	4.1 ms
		\\
		User verification&	22.5 ms
		\\
		PBFT consensus commitment  &	30 ms\\
		\hline
	\end{tabular}
	\label{table:AuthenticationCost}
\end{table}
\begin{figure}
	\centering
	\begin{tikzpicture}
	\begin{axis}[legend pos=north west, legend cell align={left}, grid=major,legend style={font=\fontsize{7}{7}\selectfont},
	xlabel=The number of healthcare users, ymax=1000,
	ylabel=Authentication time (ms), every axis plot/.append style={ultra thick}]
	\addplot [mark =square*, blue] coordinates {
		(2,240)
		(4,270)
		(6,310)
		(8,350)
		(10,400)
		(12,460)	
	};
	\addplot [mark =triangle*, red] coordinates {
		(2,310)
		(4,402)
		(6,520)
		(8,620)
		(10,720)
		(12,822)		
	};
	\legend{Centralized authentication scheme \cite{30},Our scheme with decentralized authentication}	
	\end{axis}
	\end{tikzpicture}
	\caption{Authentication latency with different numbers of healthcare users.} 
	\label{table:AuthenticationLatency}
	\vspace{-0.2in}
\end{figure}
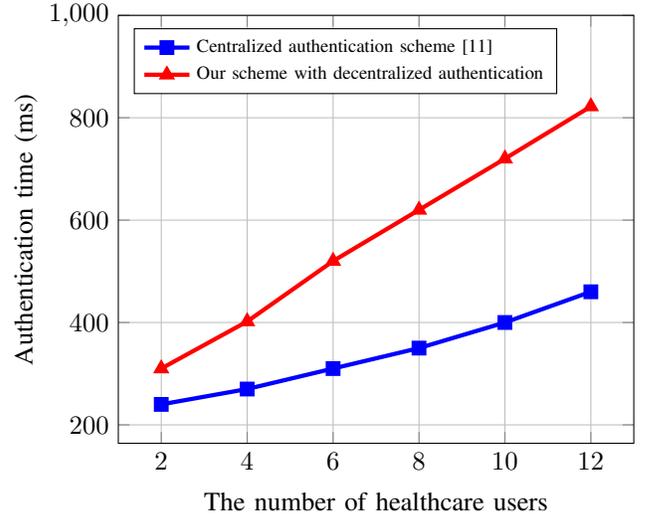

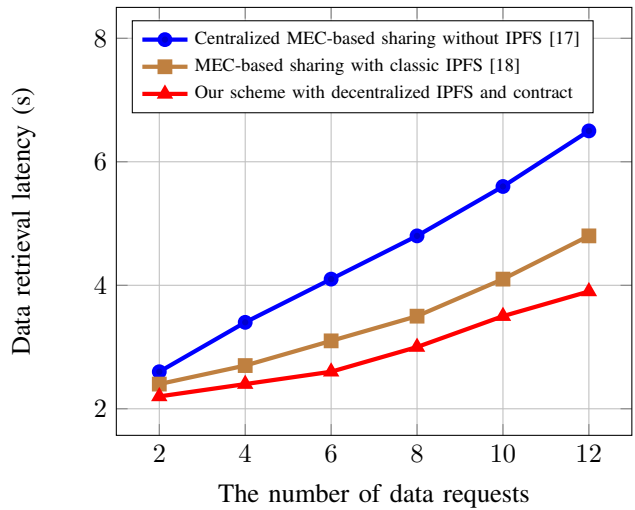
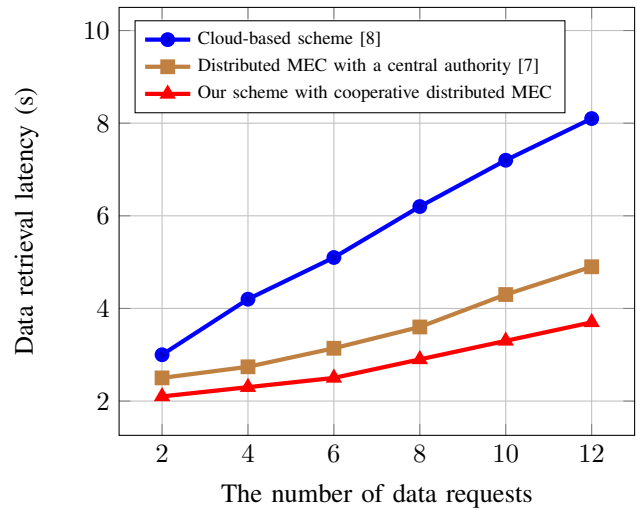
\begin{figure*}
	\centering
	\subfigure[Data retrieval latency under blockchain designs.]{
		\begin{tikzpicture}
		\begin{axis}[legend pos=north west, legend cell align={left}, grid=major,legend style={font=\fontsize{7}{7}\selectfont},
		xlabel=The number of data requests, ymax=8.5,
		ylabel=Data retrieval latency (s), every axis plot/.append style={ultra thick}]
		\addplot coordinates {
			(2,2.6)
			(4,3.4)
			(6,4.1)
			(8,4.8)
			(10,5.6)
			(12,6.5)	
		};
		\addplot [mark =square*, brown] coordinates {
			(2,2.4)
			(4,2.7)
			(6,3.1)
			(8,3.5)
			(10,4.1)
			(12,4.8)	
		};
		\addplot  [mark =triangle*, red] coordinates {
			(2,2.2)
			(4,2.4)
			(6,2.6)
			(8,3.0)
			(10,3.5)
			(12,3.9)		
		};
		\legend{Centralized MEC-based sharing without IPFS \cite{31},MEC-based sharing with classic IPFS \cite{32}, Our scheme with decentralized IPFS and contract }	
		\end{axis}
		\end{tikzpicture}}
	\subfigure[Data retrieval latency under network designs.]{
		\begin{tikzpicture}
		\begin{axis}[legend pos=north west, legend cell align={left}, grid=major,legend style={font=\fontsize{7}{7}\selectfont},
		xlabel=The number of data requests, ymax=10.5,
		ylabel=Data retrieval latency (s), every axis plot/.append style={ultra thick}]
		\addplot coordinates {
			(2,3)
			(4,4.2)
			(6,5.1)
			(8,6.2)
			(10,7.2)
			(12,8.1)	
		};
		\addplot [mark =square*, brown] coordinates {
			(2,2.5)
			(4,2.74)
			(6,3.14)
			(8,3.6)
			(10,4.3)
			(12,4.9)	
		};
		\addplot [mark =triangle*, red] coordinates {
			(2,2.1)
			(4,2.3)
			(6,2.5)
			(8,2.9)
			(10,3.3)
			(12,3.7)	
		};
		\legend{Cloud-based scheme \cite{11},Distributed MEC with a central authority \cite{5}, Our scheme with cooperative distributed MEC}	
		\end{axis}
		\end{tikzpicture}}
	
	\caption{\textcolor{black}{Data retrieval rate under different numbers of data requests.}} 
	\label{Fig:DataRetrieval}
	\vspace{-0.15in}
\end{figure*}
\subsection{Evaluation of Data Sharing Performance}
For data sharing, we investigated the authentication cost, data retrieval latency, request acceptance probability, and blockchain performance.
\subsubsection{Authentication Cost} We calculate the computation cost for the authentication process. Here, a user leverages his smartphone to submit a request with 160 bit to its MEC server for data retrieval in the blockchain network. The MEC server performs some functions such as encryption, decryption, transaction decoding for user verification. As shown in Table~\ref{table:AuthenticationCost}, the computation costs are small, and the latency of PBFT consensus is acceptable and thus suitable for time-sensitive healthcare applications. 

Besides, we also compare the authentication latency of our proposed scheme with a centralized scheme \cite{30} with the different number of healthcare users. In the proposed scheme, we organize the access authentication at the network edge where each MEC server authenticates its users by the distributed smart contract. Meanwhile, the scheme \cite{30} relies on a central authority to implement its user authentication. As shown in Fig.~\ref{table:AuthenticationLatency}, our scheme exhibits a lower latency compared to the baseline \cite{30}. This is because that the use of decentralized smart contract enables fast authentication at the network edge without passing a remote authority, which thus reduces communication overhead in the authentication process. 

\subsubsection{Data Retrieval Latency}
We investigated the data retrieval latency of our proposed model from blockchain design and network design perspectives as shown in Fig.~\ref{Fig:DataRetrieval}. We used smartphones to send data requests continuously to the MEC servers to record the results. In terms of blockchain design, we use two existing works for comparison. The first one is a centralized edge-based health sharing scheme without IPFS \cite{31} which used a centralized MEC server to serve a large hospital network and health data was stored in a classic database. The second one is an edge blockchain health sharing scheme with classic IPFS \cite{32} that integrated blockchain and edge computing without IPFS design improvement. 

From Fig.~\ref{Fig:DataRetrieval}(a), we can see that when the number of requests increases, the baseline \cite{31} has the highest data retrieval latency due to the queuing latency in the centralized MEC server. The baseline \cite{32} used a classic IPFS storage with a global DHT look-up solution that results in unnecessary communication overhead. By contrast, our scheme provides a fully decentralized solution with distributed MEC and smart contracts, which allows to implement request verification and data look-up at the network edge without using the global DHT. As a result, our scheme can achieve a minimal data retrieval latency. 

Next, we evaluated the data retrieval latency from a network design perspective as shown in Fig.~\ref{Fig:DataRetrieval}(b). We leveraged a cloud-based scheme \cite{11} and a distributed MEC with a central authority \cite{5} as the baselines for comparison. For cloud computing implementation, we employed Amazon cloud services to communicate with smartphones \cite{14}. The results in Fig.~\ref{Fig:DataRetrieval}(b) clearly show a significant improvement in our decentralized scheme with a much lower retrieval latency. This is because our scheme combines MEC, distributed smart contracts, and decentralized IPFS for fast data retrieval, without passing any external authority during the data sharing. Meanwhile, the work in \cite{11} relies on a remote cloud model which remains high latency due to excessive communication overhead. Also, the work in \cite{5} used a central authority for request verification that consumes a certain overhead for communication between the MEC servers and the authority in the request verification. The above experiment results demonstrate a lower retrieval latency cost of our approach in comparison with the existing works.

\subsubsection{Request Acceptance Probability} 
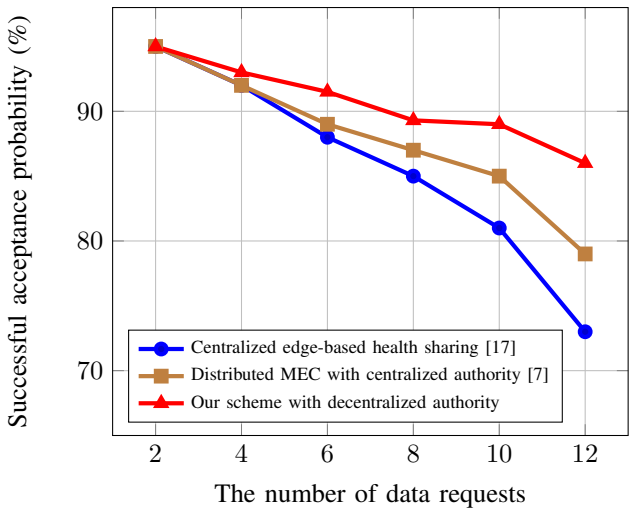
\begin{figure}
	\centering
	
	\begin{tikzpicture}
	\begin{axis}[legend pos=south west, legend cell align={left}, grid=major, legend style={font=\fontsize{7}{7}\selectfont},
	xlabel=The number of data requests, ymin=65,
	ylabel=Successful acceptance probability (\%), every axis plot/.append style={ultra thick}]
	\addplot coordinates {
		(2,95)
		(4,92)
		(6,88)
		(8,85)
		(10,81)
		(12,73)
	};
	\addplot [mark =square*, brown] coordinates {
		(2,95)
		(4,92)
		(6,89)
		(8,87)
		(10,85)
		(12,79)	
	};
	\addplot [mark =triangle*, red] coordinates {
		(2,95)
		(4,93)
		(6,91.5)
		(8,89.3)
		(10,89)
		(12,86)		
	};
	\legend{Centralized edge-based health sharing \cite{31},Distributed MEC with centralized authority \cite{5}, Our scheme with decentralized authority }	
	\end{axis}
	\end{tikzpicture}
	
	\caption{\textcolor{black}{Request acceptance probability under different numbers of data requests.}}
	\label{Fig:RequestProbability}
	\vspace{-0.1in}
\end{figure}

We added a time threshold $ \delta_u$ in the request transaction $ T_{{req}_u} $ to set up a new transaction involving the latency condition: $T_{{req}_u} \leftarrow (PKU_u||ID_u||P_{addr}||\delta_u ||timestamp)$. This threshold~represents the maximum latency that a request needs to be responded by returning the data to the requestor before the $ \delta_u$ deadline (successful request); otherwise, the request is regarded as a failed one. Here we introduce an acceptance probability function which is defined as the number of successful requests per the total requests. As shown in Fig.~\ref{Fig:RequestProbability}, the request acceptance probability of our scheme is higher than that of the other baselines \cite{31} and \cite{5}. Although the probability reduces when the number of requests increases due to the longer queue time, our decentralized scheme still yields the best performance. This can be explained by the significant processing time saving achieved in our scheme thanks to an optimized edge computing and decentralized smart contract design. 

\subsubsection{{Blockchain Performance}}
\begin{figure}
	\centering
	
	\begin{tikzpicture}
	\begin{axis}[ybar,legend pos=north west, legend cell align={left},
	ylabel=Transaction latency (ms),ymajorgrids,
	xlabel = The number of transactions,
	symbolic x coords={1, 20, 40, 60, 80},
	xtick=data,ymin=0
	]
	\addplot [fill=green!60] coordinates {(1,150) (20,460)   (40,880) (60,1370) (80,1910)  };
	
	\addplot [fill=blue!55] coordinates {(1,130) (20,222)   (40,665) (60,820) (80,1205)   };
	\legend{Ethereum blockchain \cite{32}, Our Hyperledger blockchain }
	\end{axis}
	\end{tikzpicture}
	
	\caption{Comparison on transaction latency of Hyperledger and Ethereum blockchains.} 
	\label{Fig:TransactionLatency}
	\vspace{-0.1in}
\end{figure}
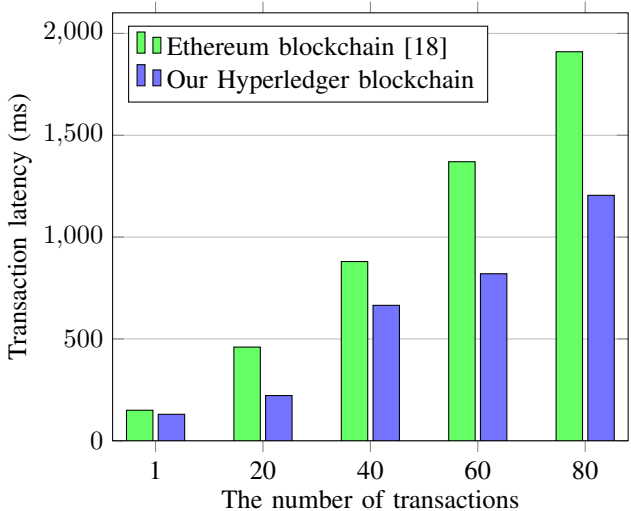

Next, we evaluate the performance of our permissioned Hyperledger Fabric blockchain and compare it with the popular permissionless Ethereum blockchain used in \cite{32}. We run the ACSC contract on three computers and use smartphones to send transaction continuously to the computers to measure the average transaction latency in the local blockchain. As indicated in Fig.~\ref{Fig:TransactionLatency}, our proposed Hyperledger Fabric blockchain exhibits a much lower transaction latency, compared to the Ethereum blockchain under the varying number of transactions. This experiment result shows the suitability of using Hyperledger Fabric blockchain for time-sensitive healthcare applications like our scenario. 

\section{Security Analysis and Discussions}
\label{Security}
\subsection{Security Analysis}
\textcolor{black}{In this sub-section, we present the attack model and then analyze theoretically some key security features acheived by our blockchain design.}
\subsubsection{Attack Model}
We consider two types of attacks: internal attack and external attack.

- \textit{Internal attack:} According to our design description in the previous sections, the MEC servers and healthcare users in this paper are considered semi-trusted in the data sharing, while the MDs are regarded as fully trusted entities in the private offloading. 
Under the semi-trusted assumption, the MEC servers would be honest but curious about the health data and thus can infer some sensitive information from transactions on the blockchain. 

- \textit{External attack:} During data offloading and sharing, external attackers can gain access to obtain health information. For example, an adversary can attack the MEC server to obtain patient information offloaded from MDs. Also, the data sharing may be vulnerable due to the data threats on the communication between the MEC server and healthcare users. 

Next, we present the main security properties of our proposed \textit{BEdgeHealth} scheme, and explain how these properties make our scheme resilient to security threats. 
\subsubsection{Security Analysis}
Our scheme is able to provide four important security features, including data privacy, authentication, traceability, and confidentiality. 

\textit{2.1) Data privacy:} Our scheme can preserve data privacy in both data offloading and data sharing schemes. In data offloading, the health data is actually encrypted by an AES encryption function in MDs when performing the offloading. Moreover, the health data offloaded from MDs is encrypted by using the private key of the MEC server $\left( C^{enc}_j \leftarrow Enc(C_j,PKM_m)\right)$.  An external adversary may not decrypt the data due to the lack of MEC server's private key. Thus, the health data privacy is preserved. In the data sharing, the data request is also encrypted $\left(T_{{enc}_u} \leftarrow Enc(T_{{req}_u},PKM_m)\right)$ so that private user information is protected against threats. Moreover, the data transactions stored in the blockchain cannot changed or modified by using immutable ledgers. Unlike recent works \cite{12, 13, 14} with a centralized IPFS on a third-party cloud which may remain data leakage, our architecture provides strong data control in the MEC network under the management of blockchain without a third party. This would eliminate single-point failures and avoid illegal data usage for better data privacy control. 

\textit{2.2) Authentication:} In our \textit{BEdgeHealth} scheme, the data sharing is authenticated in a decentralized manner with the help of the distributed ACSC contract. In our design, the smart contract works independently with the MEC server which means the authentication policies cannot be controlled by the misbehaving MEC server but managed by the global blockchain network. Any contract updates are reflected on the blockchain network and monitored by all MEC servers. This would avoid the risks of contract modifications caused by internal attacks and hence ensure reliable contract operations. In the data sharing, the data retrieval is executed only if the user information $PKU_u, ID_u$ is authenticated by the contract.

\textit{2.3) Traceability:} This is ensured by the fact that our \textit{BEdgeHealth} framework is running on a blockchain network where any data access events and user behaviours are traced by all entities \textcolor{black}{that cannot achieved in the existing works \cite{5,11,30}}. A user request  to our \textit{BEdgeHealth} system is registered by an MEC server and broadcast to all other entities in both global blockchain and local blockchains. As a result, all MEC servers and users have acknowledged a data access event when it occurs. Furthermore, one can easily trace the origin where data is modified or updated through transaction logs. \textcolor{black}{Moreover, we store health data on IPFS associated with smart contracts, instead of in the MEC server's hard drive to achieve traceability over data usage. To be clear, the raw data is stored in the IPFS while the hash value of data is kept in the smart contract. Hence, any modification or change behaviours on a data record will lead to a change in its hash value, and thus the smart contract can detect for prevention. Moreover, since all activities at an MEC server are recorded and synchronized on the global blockchain, other MEC servers can also trace them in a transparent manner. }

\textit{2.4) Confidentiality:} In \textit{BEdgeHealth}, the confidentiality of health communications is guaranteed by exploiting the standard cryptographic primitives. We employ key-based encryption solutions coupled with digital signatures for our offloading and sharing schemes. An external attack cannot gain access to the health communication due to the lack of private keys of entities (i.e., MEC servers or health users). Even if an external adversary tries to modify or change the communication protocol, the message digest on blockchain would detect such malicious actions. In fact, to modify the metadata on blockchain, an attack needs an extensive computation capability to gain the power control from all MEC servers, which may be nearly impossible to achieve in practice \cite{3}. Hence, the user confidentiality can be preserved.

Based on the above implementation results and discussions, we summarize the key features of our proposed \textit{BEdgeHealth} architecture and compare with the related works in Table~\ref{table:FeatureComparisons}. The comparison results demonstrate the advantages of our scheme over the conventional schemes, showing the usability of our \textit{BEdgeHealth} architecture in practical e-health applications. 
\subsection{Discussions}
In this article, we present a decentralized \textit{BEdgeHealth} architecture in hospital networks with a data sharing scheme and a data offloading scheme, and their performances are also verified via real-world experiments. By using MEC, our scheme is able to provide efficient health data offloading services at the network edge, proximity to MDs and IoMT networks, for user QoE improvement. Moreover, the health data can be stored and shared over the distributed hospital networks using blockchain without the need of central authority. We incorporate smart contracts with IPFS for enabling reliable access authentication and fast data retrieval in order to facilitate the data exchange among healthcare users. \textcolor{black}{However, the integration of smart contracts in IPFS possibly introduces new challenges, including security vulnerabilities which can include timestamp dependence, mishandled exceptions, reentrancy attacks on smart contracts \cite{221,222}, and fake authorization caused by misbehaving MEC servers. These security issues may result in privacy leakage or system logic modifications in the data storage process on IPFS; and thus more research efforts should be made for the full realization of smart contract-based IPFS in health data sharing applications. For example, developing an incentive mechanism may be very useful to solve fake authentication issues \cite{challenge4}. More specifically, an incentive scheme can be designed and deployed across the MEC network, aiming to incentivize the MEC server with efficient authentication and punish the MEC server with misbehaving authentication. In this way, we can mitigate the possibility of fake authentication among MEC servers.}

At present, the health data offloading scheme in our \textit{BEdgeHealth} architecture has been launched commercially at Royal Victorian Eye and Ear Hospital, Melbourne city, Australia for supporting the clinical assessments of cerebellar disease \cite{33}. This project aims to use Biokin sensors \cite{14} associated with a Biokin mobile app to collect real-world health data from patients diagnosed with cerebellar disease and then offload to the MEC server for medical analysis. Besides, our health data sharing with blockchain has been implemented in the testbed environment at Networked Sensing and Control (NSC) lab, Deakin University on the Hyperledger blockchain platform. We set up MEC servers located in different hospitals in Melbourne city and connect with an MEC server at our university, aiming to implement a health sharing among doctors working in cerebellar disease. The implementation results achieved in this work demonstrate the practicality and feasibility of our proposed \textit{BEdgeHealth} model in real-world health applications.

\begin{table}
	\scriptsize
	\centering
	\captionsetup{font=scriptsize}
	\caption{{\textcolor{black}{Comparison of security features between our proposed scheme and the existing works.}}}
	\begin{tabular}{|p{1.6cm}||c |c| c| c| c| c|}
		\hline
		\multirow{2}{*}{\textbf{Features}} &
		\multicolumn{6}{c|}{\textbf{Schemes}} \\
		&\cite{5}&	\cite{11}&	\cite{30}&	\cite{31}&	\cite{32}&	Our scheme \\
		\hline
		Decentralization&	&\checkmark	&\checkmark	&\checkmark&	\checkmark&	\checkmark \\ \hline
		Data privacy&	\checkmark&	\checkmark&	\checkmark	&\checkmark&	\checkmark&	\checkmark \\ \hline
		Decentralized storage &		&	&	&&	\checkmark&	\checkmark\\ \hline
		Authentication	&\checkmark&	\checkmark&	\checkmark&	\checkmark&	&	\checkmark\\ \hline
		Traceability &	&	&	&	\checkmark&	\checkmark	&\checkmark\\\hline
		Confidentiality	&\checkmark&	\checkmark	&\checkmark&	\checkmark&	\checkmark&	\checkmark\\ \hline
		Offloading and sharing services &	&	&	&	&	&	\checkmark \\
		\hline
	\end{tabular}
	\label{table:FeatureComparisons}
	\vspace{-0.1in}
\end{table}

\section{Conclusions and Future Work}
\label{Conclude}
This paper has proposed a new decentralized health architecture, called \textit{BEdgeHealth} that employs MEC and blockchain for health data offloading and sharing in distributed hospital networks. We have first proposed a privacy-aware data offloading scheme where MDs can offload IoMT health data to the nearby MEC server under system constraints. Then, a new data sharing scheme is introduced by using blockchain and smart contracts to enable secure data exchange among healthcare users in different hospitals. To realize access management, we have developed an ACSC contract that enables decentralized user authentication at the network edge without requiring central authority, which would ensure authentication reliability and reduce network latency. We have implemented various real-world experiments to verify the effectiveness of the proposed \textit{BEdgeHealth} architecture. The implementation results have demonstrated the significant advantages of the proposed offloading scheme over the other baseline methods in terms of reduced time latency, energy consumption, and better memory usage. Moreover, the data sharing scheme can achieve fast data retrieval with improved blockchain performance, compared to the existing works. The evaluations also prove the high system security of our design, showing the feasibility of the proposed model for healthcare applications.

\textcolor{black}{Future work is in progress to extend our blockchain-MEC model to many other healthcare systems, including speech and video data management and secure real-time health monitoring systems.}

\bibliography{Ref}

\begin{thebibliography}{10}
\providecommand{\url}[1]{#1}
\csname url@samestyle\endcsname
\providecommand{\newblock}{\relax}
\providecommand{\bibinfo}[2]{#2}
\providecommand{\BIBentrySTDinterwordspacing}{\spaceskip=0pt\relax}
\providecommand{\BIBentryALTinterwordstretchfactor}{4}
\providecommand{\BIBentryALTinterwordspacing}{\spaceskip=\fontdimen2\font plus
\BIBentryALTinterwordstretchfactor\fontdimen3\font minus
  \fontdimen4\font\relax}
\providecommand{\BIBforeignlanguage}[2]{{%
\expandafter\ifx\csname l@#1\endcsname\relax
\typeout{** WARNING: IEEEtran.bst: No hyphenation pattern has been}%
\typeout{** loaded for the language `#1'. Using the pattern for}%
\typeout{** the default language instead.}%
\else
\language=\csname l@#1\endcsname
\fi
#2}}
\providecommand{\BIBdecl}{\relax}
\BIBdecl

\bibitem{DinhGlobe}
D.~C. Nguyen, P.~N. Pathirana, M.~Ding, and A.~Seneviratne, ``Blockchain and
  edge computing for decentralized {EMRs} sharing in federated healthcare,'' in
  \emph{2020 IEEE Global Communications Conference (GLOBECOM)}, 2020.

\bibitem{3}
M.~A. Rahman, M.~S. Hossain, G.~Loukas, E.~Hassanain, S.~S. Rahman, M.~F.
  Alhamid, and M.~Guizani, ``Blockchain-based mobile edge computing framework
  for secure therapy applications,'' \emph{IEEE Access}, vol.~6, pp.
  72\,469--72\,478, 2018.

\bibitem{2}
P.~Verma and S.~K. Sood, ``Fog assisted-{IoT} enabled patient health monitoring
  in smart homes,'' \emph{IEEE Internet of Things Journal}, vol.~5, no.~3, pp.
  1789--1796, 2018.

\bibitem{mcghin2019blockchain}
T.~McGhin, K.-K.~R. Choo, C.~Z. Liu, and D.~He, ``Blockchain in healthcare
  applications: Research challenges and opportunities,'' \emph{Journal of
  Network and Computer Applications}, 2019.

\bibitem{yazdinejad2020decentralized}
A.~Yazdinejad, G.~Srivastava, R.~M. Parizi, A.~Dehghantanha, K.-K.~R. Choo, and
  M.~Aledhari, ``Decentralized authentication of distributed patients in
  hospital networks using blockchain,'' \emph{IEEE Journal of Biomedical and
  Health Informatics}, 2020.

\bibitem{4}
M.~Asif-Ur-Rahman, F.~Afsana, M.~Mahmud, M.~S. Kaiser, M.~R. Ahmed,
  O.~Kaiwartya, and A.~James-Taylor, ``Toward a heterogeneous mist, fog, and
  cloud-based framework for the internet of healthcare things,'' \emph{IEEE
  Internet of Things Journal}, vol.~6, no.~3, pp. 4049--4062, 2018.

\bibitem{5}
R.~Saha, G.~Kumar, M.~K. Rai, R.~Thomas, and S.-J. Lim, ``Privacy ensured
  e-healthcare for fog-enhanced {IoT} based applications,'' \emph{IEEE Access},
  vol.~7, pp. 44\,536--44\,543, 2019.

\bibitem{11}
J.~Liu, X.~Li, L.~Ye, H.~Zhang, X.~Du, and M.~Guizani, ``{BPDS}: A blockchain
  based privacy-preserving data sharing for electronic medical records,'' in
  \emph{2018 IEEE Global Communications Conference (GLOBECOM)}, 2018, pp. 1--6.

\bibitem{13}
S.~Wang, Y.~Zhang, and Y.~Zhang, ``A blockchain-based framework for data
  sharing with fine-grained access control in decentralized storage systems,''
  \emph{IEEE Access}, vol.~6, pp. 38\,437--38\,450, 2018.

\bibitem{14}
D.~C. Nguyen, P.~N. Pathirana, M.~Ding, and A.~Seneviratne, ``Blockchain for
  secure {EHRs} sharing of mobile cloud based e-health systems,'' \emph{IEEE
  Access}, vol.~7, pp. 66\,792--66\,806, 2019.

\bibitem{30}
S.~Jiang, H.~Wu, and L.~Wang, ``Patients-controlled secure and
  privacy-preserving {EHRs} sharing scheme based on consortium blockchain,'' in
  \emph{2019 IEEE Global Communications Conference (GLOBECOM)}, 2019, pp. 1--6.

\bibitem{8}
R.~M. Abdelmoneem, A.~Benslimane, E.~Shaaban, S.~Abdelhamid, and S.~Ghoneim,
  ``A cloud-fog based architecture for {IoT} applications dedicated to
  healthcare,'' in \emph{IEEE International Conference on Communications
  (ICC)}, 2019, pp. 1--6.

\bibitem{9}
D.~Giri, M.~S. Obaidat, and T.~Maitra, ``Sechealth: An efficient fog based
  sender initiated secure data transmission of healthcare sensors for e-medical
  system,'' in \emph{IEEE Global Communications Conference}, 2017, pp. 1--6.

\bibitem{10}
M.~Min, X.~Wan, L.~Xiao, Y.~Chen, M.~Xia, D.~Wu, and H.~Dai, ``Learning-based
  privacy-aware offloading for healthcare {IoT} with energy harvesting,''
  \emph{IEEE Internet of Things Journal}, vol.~6, no.~3, pp. 4307--4316, 2018.

\bibitem{12}
H.~Guo, W.~Li, M.~Nejad, and C.-C. Shen, ``Access control for electronic health
  records with hybrid blockchain-edge architecture,'' in \emph{2019 IEEE
  International Conference on Blockchain}, 2019, pp. 44--51.

\bibitem{add1}
H.~H. Elazhary and S.~F. Sabbeh, ``The {W5} framework for computation
  offloading in the internet of things,'' \emph{IEEE Access}, vol.~6, pp.
  23\,883--23\,895, 2018.

\bibitem{31}
X.~Li, X.~Huang, C.~Li, R.~Yu, and L.~Shu, ``Edgecare: leveraging edge
  computing for collaborative data management in mobile healthcare systems,''
  \emph{IEEE Access}, vol.~7, pp. 22\,011--22\,025, 2019.

\bibitem{32}
P.~C.~M. Arachchige, P.~Bertok, I.~Khalil, D.~Liu, S.~Camtepe, and
  M.~Atiquzzaman, ``A trustworthy privacy preserving framework for machine
  learning in industrial {IoT} systems,'' \emph{IEEE Transactions on Industrial
  Informatics}, 2020.

\bibitem{25}
R.~Kumar, N.~Marchang, and R.~Tripathi, ``Distributed off-chain storage of
  patient diagnostic reports in healthcare system using {IPFS} and
  blockchain,'' in \emph{2020 International Conference on COMmunication Systems
  \& NETworkS (COMSNETS)}, 2020, pp. 1--5.

\bibitem{26}
S.~Pongnumkul, C.~Siripanpornchana, and S.~Thajchayapong, ``Performance
  analysis of private blockchain platforms in varying workloads,'' in
  \emph{26th International Conference on Computer Communication and Networks
  (ICCCN)}, 2017, pp. 1--6.

\bibitem{banerjee2014monitoring}
T.~Banerjee, M.~Enayati, J.~M. Keller, M.~Skubic, M.~Popescu, and M.~Rantz,
  ``Monitoring patients in hospital beds using unobtrusive depth sensors,'' in
  \emph{36th Annual International Conference of the IEEE Engineering in
  Medicine and Biology Society}, 2014, pp. 5904--5907.

\bibitem{16}
D.~C. Nguyen, P.~N. Pathirana, M.~Ding, and A.~Seneviratne, ``Privacy-preserved
  task offloading in mobile blockchain with deep reinforcement learning,''
  \emph{IEEE Transactions on Network and Service Management}, pp. 1--1, 2020.

\bibitem{18}
I.~Elgendy, W.~Zhang, C.~Liu, and C.-H. Hsu, ``An efficient and secured
  framework for mobile cloud computing,'' \emph{IEEE Transactions on Cloud
  Computing}, pp. 1--1, 2018.

\bibitem{kumar2017comparative}
B.~S. Kumar, V.~R. Raj, and A.~Nair, ``Comparative study on {AES} and {RSA}
  algorithm for medical images,'' in \emph{2017 International Conference on
  Communication and Signal Processing (ICCSP)}, 2017, pp. 0501--0504.

\bibitem{19}
K.~Tamilarasi and A.~Jawahar, ``Medical data security for healthcare
  applications using hybrid lightweight encryption and swarm optimization
  algorithm,'' \emph{Wireless Personal Communications}, pp. 1--22, 2020.

\bibitem{elkady2016modified}
S.~K. ElKady and H.~M. Abdelsalam, ``A modified particle swarm optimization
  algorithm for solving capacitated maximal covering location problem in
  healthcare systems,'' in \emph{Applications of Intelligent Optimization in
  Biology and Medicine}, 2016, pp. 117--133.

\bibitem{27}
\BIBentryALTinterwordspacing
``Hyperledger fabric,'' Accessed Jun. 2020. [Online]. Available:
  \url{https://www.hyperledger.org/use/fabric}
\BIBentrySTDinterwordspacing

\bibitem{28}
\BIBentryALTinterwordspacing
``Docker software,'' Accessed Jun. 2020. [Online]. Available:
  \url{https://www.docker.com/get-started}
\BIBentrySTDinterwordspacing

\bibitem{29}
\BIBentryALTinterwordspacing
``Interplanetary file system (ipfs),'' Accessed Jun. 2020. [Online]. Available:
  \url{https://docs.ipfs.io/}
\BIBentrySTDinterwordspacing

\bibitem{221}
S.~Rouhani and R.~Deters, ``Security, performance, and applications of smart
  contracts: A systematic survey,'' \emph{IEEE Access}, vol.~7, pp.
  50\,759--50\,779, 2019.

\bibitem{222}
M.~Wohrer and U.~Zdun, ``Smart contracts: security patterns in the ethereum
  ecosystem and solidity,'' in \emph{2018 International Workshop on Blockchain
  Oriented Software Engineering (IWBOSE)}, 2018, pp. 2--8.

\bibitem{challenge4}
E.~K. Wang, Z.~Liang, C.-M. Chen, S.~Kumari, and M.~K. Khan,
  ``\BIBforeignlanguage{en}{{PoRX}: {A} {Reputation} {Incentive} {Scheme} for
  {Blockchain} {Consensus} of {IIoT}},'' \emph{\BIBforeignlanguage{en}{Future
  Generation Computer Systems}}, vol. 102, pp. 140--151, Jan. 2020.

\bibitem{33}
D.~Phan, N.~Nguyen, P.~N. Pathirana, M.~Horne, L.~Power, and D.~Szmulewicz, ``A
  random forest approach for quantifying gait ataxia with truncal and
  peripheral measurements using multiple wearable sensors,'' \emph{IEEE Sensors
  Journal}, vol.~20, no.~2, pp. 723--734, 2019.

\end{thebibliography}
\bibliographystyle{IEEEtran}

\end{document}